\documentclass[a4paper,11pt]{article}
\usepackage[latin1]{inputenc}
\usepackage{graphicx}
\usepackage[T1]{fontenc}
\usepackage[english]{babel}
\usepackage{listings}
\usepackage{amssymb}
\usepackage{color}
\usepackage{xcolor}
\usepackage{mathrsfs}
\usepackage[position=top]{caption}
\usepackage[captionskip=0pt,topadjust=-10pt]{subfig}
\usepackage{enumerate}
\usepackage{rotating}
\usepackage{setspace}
\long\def\symbolfootnote[#1]#2{\begingroup%
\def\thefootnote{\fnsymbol{footnote}}\footnote[#1]{#2}\endgroup}
\usepackage[hypertexnames=false]{hyperref}
\usepackage{amsmath}
\usepackage{fancyhdr}
\usepackage{verbatim,lineno}
\usepackage[longnamesfirst]{natbib}

\newcommand{\MCMC}{Markov chain Monte Carlo{}}
\newcommand{\factormsv}{factor multivariate stochastic volatility{}}
\newcommand{\fmsv}{factor MSV{}}
\newcommand{\Fmsv}{Factor MSV{}}
\newcommand{\MH}{Metropolis-Hastings{}}
\newcommand{\kstep}{\lq $k$-step iteration\rq{}}
\newcommand{\optim}{\lq optimization\rq{}}
\newcommand{\dr}{\lq delayed rejection\rq{}}
\newcommand{\usv}{univariate SV{}}
\newcommand{\arwm}{adaptive random walk Metropolis{}}
\newcommand{\mcem}{Monte Carlo EM{}}
\newcommand{\fmgarch}{factor MGARCH}
\newcommand{\ugarch}{univariate GARCH}

\newcommand{\btheta}{\mbox{\boldmath $\theta$}}
\newcommand{\bTheta}{\mbox{\boldmath $\Theta$}}
\newcommand{\bzeta}{\mbox{\boldmath $\zeta$}}
\newcommand{\bEta}{\mbox{\boldmath $\eta$}}
\newcommand{\by}{\mbox{\boldmath $y$}}
\newcommand{\bz}{\mbox{\boldmath $z$}}
\newcommand{\bh}{\mbox{\boldmath $h$}}
\newcommand{\bK}{\mbox{\boldmath $K$}}
\newcommand{\bF}{\mbox{\boldmath $f$}}
\newcommand{\bb}{\mbox{\boldmath $b$}}
\newcommand{\bB}{\mbox{\boldmath $B$}}
\newcommand{\bC}{\mbox{\boldmath $C$}}
\newcommand{\bS}{\mbox{\boldmath $S$}}
\newcommand{\bV}{\mbox{\boldmath $V$}}

\newcommand{\be}{\mbox{\boldmath $e$}}

\newcommand{\bmf}{\mbox{\boldmath $f$}}
\newcommand{\bu}{\mbox{\boldmath $u$}}
\newcommand{\bI}{\mbox{\boldmath $I$}}

\newcommand{\ttheta}{{\widetilde \theta}}
\newcommand{\bbeta}{\mbox{\boldmath $\beta$}}
\newcommand{\betahat}{{\widehat  \bbeta}}
\newcommand{\bttheta}{\mbox{\boldmath{$\ttheta$}}}
\newcommand{\thetahat}{\mbox{\boldmath{$\widehat \theta$}}}
\newcommand{\Sigmahat}{\mbox{\boldmath{$\widehat \Sigma$}}}
\newcommand{\bSigma}{\mbox{\boldmath $\Sigma$}}
\newcommand{\rmIG}{\ensuremath{\textrm{IG}}}
\newcommand{\half}{\frac12}
\newcommand{\MVN}{\text{MVN}}

\newcommand{\bthetaold}{\btheta^\mathrm{old}}
\newcommand{\rhohat}{\widehat \rho}

\AtBeginDocument{}

\begin{document}
\setpagewiselinenumbers
\Large
\title{Computationally Efficient Estimation of
 Factor Multivariate Stochastic Volatility Models}
\author{Weijun Xu \hspace{0.15cm} Li Yang \hspace{0.15cm} Robert Kohn \\
  University of New South Wales, Sydney}
\date{\today}
\maketitle

\begin{abstract}
\large
An \MCMC{} simulation method based on a two stage delayed rejection \MH{} algorithm is proposed to estimate a \factormsv{} model.
The first stage uses \kstep{} towards the mode, with $k$ small, and the second stage uses an adaptive random walk proposal density. The marginal likelihood approach of \cite{chib:1995}
is used to choose the number of factors, with the posterior density ordinates approximated by Gaussian copula. Simulation and real data applications suggest that the proposed simulation method is computationally much more efficient than the approach of
\cite{chib:nardari:shephard:2006}. This increase in computational efficiency is particularly important in calculating marginal likelihoods because it is necessary to carry out the simulation a number of times to estimate the posterior ordinates for a given marginal likelihood. In addition to the \MCMC{} method, the paper also proposes a fast approximate EM method to estimate the \factormsv{}. The estimates from the approximate EM method are of interest in their own right but are especially useful as initial inputs to \MCMC{} methods, making them more efficient computationally. The methodology is illustrated using simulated and real examples.
\end{abstract}

\thispagestyle{empty}
\clearpage

\section{Introduction} \label{sec:intr}
Factor multivariate stochastic volatility (\fmsv) models are increasingly used in the
financial economics literature because they can  model the volatility dynamics
of a large system of financial and economic time series when the common features in
these series can be captured by a small number of latent factors.  These models
naturally link in with factor pricing models
such as the arbitrage pricing theory (APT) model of \cite{ross:1976},  which is  built
on the existence of a set of common factors underlying all asset returns,  and the capital
asset pricing model (CAPM) of \cite{sharpe:1964} and \cite{lintner:1965} where
the \lq market return\rq{}  is the common risk factor affecting all
the assets. However, unlike  \fmsv{} models, standard factor pricing models usually do not
attempt to model the dynamics of the volatilities of the asset returns and instead
assume the second moments are constant. \Fmsv{} models have recently been applied
to important problems in financial markets such as asset
allocation \citep[e.g.][]{aguilar:west:2000,han:2006}
and asset pricing \citep[e.g.][]{nardari:scruggs:2006}.  Empirical evidence suggests
that \fmsv{} models are  a promising approach  for  modeling multivariate
time-varying volatility, explaining excess asset returns, and generating optimal
portfolio strategies.

A computationally efficient method of estimating a high dimensional dynamic \fmsv{} model
is necessary if such a model is to be applied to financial problems to make decisions
in a timely way. For example, when new information becomes available to financial markets,
fund managers need to quickly incorporate it into their portfolio strategies,
so the speed with which a \fmsv{} can be re-estimated is
an important practical issue. However, based on results reported in the literature
\citep[e.g.][]{chib:nardari:shephard:2006} and in our article, estimating a \fmsv{} using
current Bayesian simulation methods can take a considerable amount of time
when the number of assets is moderate to large which limits
the applicability of \fmsv{} models in real-time applications.
Hence, the main purpose of our article is to develop  estimation methods
for \fmsv{} models that are computationally more efficient in order to  enhance their applicability.

There are  a number of methods to estimate \fmsv{} models such as
quasi-maximum likelihood  \citep[e.g.][] {harvey:ruiz:shephard:1994},
simulated maximum likelihood
\citep[e.g.][]{liesenfeld:richard:2006,jungbacker:koopman:2006},
and Bayesian MCMC  simulation \citep[e.g.][]{chib:nardari:shephard:2006} .
For high dimensional problems,
\citep[e.g.][]{chib:nardari:shephard:2006, han:2006}, Bayesian MCMC simulation is the most efficient estimation method,
with alternative estimation methods having difficulty handling  high dimensions.

The main practical and computational advantage of  a  \fmsv{}  model
is its parsimony, where all the variances and covariances of a vector of time series
are modeled by a low-dimensional SV structure governed by common factors.
In a series of papers \cite{kim:shephard:chib:1998},
\cite{chib:nardari:shephard:2002},  \cite{chib:nardari:shephard:2006}, and
\cite{omori:2007},
Chib and Shephard and their coauthors consider a variety of univariate and
multivariate stochastic volatility (MSV) models whose error distributions range
from Gaussian to Student-t, and that allow for both symmetric and asymmetric
conditional distributions. In the  multivariate case, the correlation between
variables is governed by several latent factors.
A computationally  efficient estimation method for a  \fmsv{} model depends on how efficiently a \usv{} model is estimated and how  efficiently  the latent factors and their corresponding coefficients are estimated.  Our article first improves the computational efficiency of estimating  a \usv{} model and then extends the estimation method to
 a \fmsv{} model.

Bayesian \MCMC{} (MCMC) simulation is a convenient method to estimate a \usv{} model.  The model is transformed to a linear state space model with an error term in the observation equation having a log-chisquared distribution which is approximated by a mixture of normals \citep[e.g.][]{kim:shephard:chib:1998}.
One of the most efficient MCMC methods for estimating a \usv{} model is based on
the Metropolis-Hastings method which uses a multivariate t proposal distribution for
sampling the model parameters in blocks, with the location and scale matrix obtained
from the mode of the  posterior distribution \citep[see][]{chib:greenberg:1995}.
This method usually requires numerical optimization and is used in the papers
by Chib and Shephard and their coauthors both for univariate SV models and for factor
MSV models. We refer to this as the \optim{} method.   Although the \optim{}  method
is a powerful approach for most block-sampling schemes, it is computationally
demanding, especially in \fmsv{} models,  because it is necessary to apply it within
each MCMC iteration.

Our article provides an alternative \MH{} method for sampling the parameters in SV models,
which is more efficient than \optim{}  method, and is based on a two stage \dr{}
algorithm. The first stage consists of $k$-steps of a Newton-Raphson iteration towards
the mode of the posterior distribution, with $k$ small. We refer to this as the
\kstep{} stage. The second stage consists of an adaptive random walk Metropolis
(ARWM) proposal whose purpose is to keep the chain moving in small steps if the first stage
proposal is poor in some region of the parameter space. If only the single stage proposal is used then the Markov chain may remain stuck in such a region for many iterations.

We compare our methods with the \optim{}  method using simulated data for
\usv{} and \fmsv{} models and show that the \dr{} methods is
computationally much more efficient than the \optim{} method when both computing
time and inefficiency factors are taken into account.
The gains are more than 200\% for \usv{} models and more than 500\% for \fmsv{} models.

In addition to MCMC methods, we also consider an approximate \mcem{}
method to estimate SV models, which is much faster than the MCMC based methods,
and yields good parameter estimates. The speed advantage is  especially important
in high dimensional \fmsv{} models. An important use of the \mcem{} method is to provide
initial parameter values for the MCMC based methods, especially in high dimensions.

An important practical issue in estimating \fmsv {} models is determining the number
of latent factors.  In factor asset pricing models, a popular approach is to rely on
 intuition and theory as guides to come up with a list of observed variables as proxies
 of the unobserved theoretical factors.  The adequacy of the list of observed variables
 has crucial effects on the covariance structure of idiosyncratic shocks.  We choose the number of factors in \fmsv{} models by marginal likelihood, based on the approach of \cite{chib:1995} and with the conditional posterior ordinates approximated by Gaussian copulas that are fitted using MCMC iterates.
Estimating the marginal posterior ordinates for a given marginal likelihood usually
requires several MCMC runs which makes the greater computational efficiency of our
methods compared to that of the \optim{} method even more important.

Delayed rejection for \MH{} is proposed by \cite{mira:1999} in her thesis and
extended in \cite{green:mira:2001}.  \kstep{} is proposed by \cite{gamerman:1997}
for random effects models and implemented using iteratively reweighted least squares.
The method is refined by \cite{villani:2009} using Newton-Raphson steps and is applied
in their paper for nonparametric regression density estimation.
In the second stage of \dr{} we use the \arwm{} algorithm of
\cite{roberts:rosenthal:2009} . The copula based approach of estimating posterior
ordinates is proposed by \cite{nott:kohn:xu:fielding:2009}.

The rest of the paper is organized as follows.
Section \ref{sec:unimodel}  reviews the \lq \optim\rq{}
method and introduces our approach  for estimating \usv{} models.
Section~\ref{Sec: factor msv models} extends the
MCMC approach to \fmsv{} models and also proposes the \mcem{} approach.
Section 4 shows how to approximate the marginal likelihood in complex models using a
Gaussian copula approach. Sections 5 and 6 compare the various estimation methods
using simulated and real data.  Section 7 discusses the proposed methods for other
applications such as GARCH type models. Section 8 concludes the paper.

\section{Computational methods for univariate stochastic volatility models }
\label{sec:unimodel}
Although there are a number of \usv{} models discussed in literature
(see the papers by Chib and Shephard and their coauthors) our article only deals
with one simple model in this family, which  is the basis for the \fmsv{} models
discussed in section~\ref{Sec: factor msv models}.  The centered parametrization
version of this model \citep[see][]{pitt:shephard:1999} is
\begin{align}
\begin{split}
y_t&=\exp(h_t/2)e_t, \hspace{3cm}  e_t\sim N(0,1) \ , \\
h_t&=\mu+\phi(h_{t-1}-\mu)+\sigma\eta_t, \hspace{1.25cm} \eta_t\sim N(0,1)\ ,  \label{eq:uni}
\end{split}
\end{align}
where $y_t$ is the mean-adjusted return of an asset at time $t$ and  $h_t$ is its log volatility which itself is governed by an autoregressive (AR) process with mean $\mu$, persistence parameter $\phi$ and a Gaussian noise term with standard deviation $\sigma$. The two noise terms $e_t$ and $\eta_t$ are assumed to be uncorrelated. We assume that $|\phi|<1$ to ensure that the log volatility $h_t$ is stationary.

The \usv{}  model can be  transformed into  linear state space form by writing
\begin{align}
\begin{split}
z_t&=\log(y_t^2)=h_t+\log(e_t^2),  \hspace{1.3cm} e_t\sim N(0,1)\ , \\
h_t&=\mu+\phi(h_{t-1}-\mu)+\sigma\eta_t, \hspace{1.25cm} \eta_t\sim N(0,1)\ .
\label{eq: uni trans}
\end{split}
\end{align}
Unlike the standard Gaussian linear state space model,
the error term $\log(e_t^2)$ in the measurement equation~\eqref{eq: uni trans} is non-Gaussian so that it is not possible to estimate the model parameters by maximum likelihood estimation using the Kalman filter.
 \cite{harvey:ruiz:shephard:1994} approximate the log-chisquared distribution with one degree of freedom at  equation~\eqref{eq: uni trans} by a Gaussian distribution having the same mean and variance, but the approximation is unreliable.
In the Bayesian literature several Gibbs sampling methods are proposed \citep[e.g.][]{jacquier:polson:rossi:1994}, but as discussed in \cite{kim:shephard:chib:1998} these methods are computationally inefficient. \cite{carter:kohn:1997} and  \cite{kim:shephard:chib:1998} show that a finite mixture of normals can provide a very good approximation the log-chisquare distribution with one degree of freedom.  \cite{carter:kohn:1997} use a mixture with 5 components and \cite{kim:shephard:chib:1998} use a normal  mixture with 7 components and correct the approximation with a \MH{} step.
We write such a normal mixture approximation as
 \begin{align}\label{eq: norm approx}
p(\epsilon_t) &\approx \sum_{i}\pi_i\phi_N(z_t;m_i,\upsilon_i), \hspace{1cm} \epsilon_t=\log(e_t^2)\ ,
\end{align}
where $\phi_N(z;m,\upsilon)$ is a univariate normal density in $z$ with mean $m$ and variance $\upsilon$. The weights $\pi_i$, means $m_i$ and variances $\upsilon_i$ for each normal component are given by \cite{carter:kohn:1997} and \cite{kim:shephard:chib:1998}
 for their approximations. Our article  uses the 5-component approximation in \cite{carter:kohn:1997}.

 Using equation~\eqref{eq: norm approx}, we can write equation~\eqref{eq: uni trans}
 as a conditionally Gaussian state space model by introducing a sequence of discrete latent variables $K_t$ each  taking the five values $1, \dots, 5$ such that conditionally on $h_t$ and $K_t = i$, $z_t$ is Gaussian with mean $h_t + m_i$ and variance $\upsilon_i$, i.e. $z_t|K_t=i\sim \phi_N(z_t; h_t+m_i,\upsilon_i)$.

The sampling scheme for the \usv{} model at  equation~\eqref{eq: uni trans}
and its various extensions in the series of papers by Chib and Shephard and their coauthors  are very similar and can be summarized as follows.

\subsection{The Optimization MCMC Sampling Method}
\begin{enumerate}
\item Initialize $\btheta=(\mu,\phi,\sigma)'$ and $\bh=(h_1,...,h_T)'$.
\item Sample the indicator variables from
$\bK\sim p(\bK|\bz,\bh)$,
so that
\begin{align*}		 p(K_t=i|z_t,h_t)=\frac{\pi(K_t=i)\phi_N(z_t;h_t+m_i,\upsilon_i)}
{\sum_{j=1}^{5}\pi(K_t=j)\phi_N(z_t;h_t+m_j,\upsilon_j)}\ .
	\end{align*}
\item Jointly sample $\btheta,\bh\sim p(\btheta,\bh|\bz,\bK)$.
		\begin{enumerate}[3.1.]
		\item Sample $\btheta\sim p(\btheta|\bz,\bK)$ (for convenience,  we suppress $\bK$ in the densities below) using the Metropolis-Hastings algorithm as
\begin{enumerate}[a.]
\item Build the proposal distribution $q_T({\thetahat},\Sigmahat ,v)$
 for the target, where ${\thetahat}$ is the value of $\btheta$ that maximizes
 $\log p(\bz|\btheta)$; the density  $ p(\bz|\btheta)$ is calculated using the
  Kalman Filter
\citep[see][for details]{anderson:moore:1979}.
$\Sigmahat$ is the negative of the inverse of the Hessian matrix of the objective function evaluated at ${\thetahat}$ and $v$ is the degrees of freedom for the multivariate-t distribution;
\item Sample a candidate
value $\btheta^1\sim q_T({\thetahat},\Sigmahat,v)$;			
				\item Accept the candidate value $\btheta^1$ with probability
				\begin{align*}
\alpha(\btheta^0\rightarrow\btheta^1)=\min{\left\{1,\frac{\pi(\btheta^1)p(\bz|\btheta^1)
q_T(\btheta^0|\widehat{\btheta},\Sigmahat,v)}{\pi(\btheta^0)p(\bz|\btheta^0)
q_T(\btheta^1|\thetahat,\Sigmahat,v)}\right\}}\ ,
\end{align*}		
where $\btheta^0$ represents the current draw and $\pi(\btheta)$ is the prior. If the candidate value $\btheta^1$ is rejected, the current value $\btheta^0$ is retained as the next draw.		
\end{enumerate}	
\item Sample the latent state variable $h\sim p(\bh|\bz,\btheta,\bK)$ using a Forwards-Filtering-Backwards-Sampling (FFBS) algorithm such as \cite{carter:kohn:1994}, \cite{fruwirth-schnatter:1994} or \cite{dejong:shephard:1995}.
		\end{enumerate}
\item Go to Step 2.
\end{enumerate}

Step 3.1.a builds the proposal distribution for block sampling the parameters,
and follows the method proposed by
\cite{chib:greenberg:1995} where optimization is carried out at  each MCMC iteration to obtain the mode of the log likelihood function. This is the reason  that
we call this sampling scheme the \lq optimization\rq{}  method.
One of the main advantages of the
\lq optimization\rq{}
 method is that it jointly samples the parameters  $\btheta$ and the latent states $\bh$ , which avoids the dependence between the two in an MCMC scheme.
 Furthermore,  it samples  $\btheta =(\mu,\phi,\sigma)^\prime$ in one block which also diminishes the dependence in the sampling between the parameters. However, the computational burden of this method  is  heavy because it is necessary to do the optimization at  each MCMC iteration. This problem is more severe for large datasets since the evaluation of the log likelihood at each Newton-Raphson iteration is more time consuming.

\subsection{Improved methods for sampling the model parameters}\label{ss: improved methods}
The major computational load of the `optimization' method comes
from searching for the mode of the likelihood
$p(\bz|\btheta,\bK)$ in order to construct the proposal distribution for $\btheta$. One way to reduce the computational
burden is to reduce the dimension of the parameter space from three to two by augmenting
$\mu$ to the latent state vector $\bh$ and rewriting the \usv{} model at
equation \eqref{eq:uni} in the non-centered form
 \citep[see][]{pitt:shephard:1999} as
\begin{align}
\begin{split}
z_t  =\log(y_t^2) & =(1,1)\begin{pmatrix} h_t-\mu_t\\ \mu_t\end{pmatrix}+\log(e_t^2),    \qquad  e_t\sim N(0,1)\ , \\
 \begin{pmatrix} h_t-\mu_t\\ \mu_t\end{pmatrix} & =\begin{pmatrix}\phi & 0\\ 0 & 1\end{pmatrix}
 \begin{pmatrix} h_{t-1}-\mu_{t-1}\\ \mu_{t-1}\end{pmatrix}+ \begin{pmatrix} \sigma\\ 0\end{pmatrix}  \eta_t,    \qquad \eta_t\sim N(0,1) \label{eq:uni2}
\end{split}
\end{align}
with initial values
\begin{align*}
\begin{pmatrix}  h_1-\mu_1\\ \mu_1\end{pmatrix} \sim N\left( \begin{pmatrix} 0\\ m_\mu\end{pmatrix}, \begin{pmatrix}\sigma^2/(1-\phi^2)+V_\mu & -V_\mu\\ -V_\mu & V_\mu\end{pmatrix}\right )\ .
\end{align*}
The prior for $\mu$ is
$N(m_\mu,V_\mu)$ and the state vector at time $t$
 is  redefined as $\bh_t= (h_t - \mu, \mu)^\prime$.

The computational gain by augmenting $\mu$ to the latent state $h_t$ is not very
significant in the univariate case because optimization is still required. However, it
becomes more important in the multivariate case. In this section
we propose an alternative sampling method that builds a  proposal density
for $\btheta=(\phi,\sigma)'\sim p(\btheta|\bz,\bK)$  but does not require optimization.

\subsubsection{\kstep{} sampling method}
Unlike the `optimization' MCMC method where it is necessary to find the mode before building the proposal distribution, the
\kstep{} method builds the  proposal distribution using $k$ Newton-Raphson iterations towards the mode of
$L(\btheta) = \log p(\bz\mid  \btheta, \bK) $, with $k$ small, typically $k = 1$ or 2.
For notational convenience we omit to indicate dependence on the indicators $K$ below.
The iterations based on the second order Taylor series expansion of $L(\btheta)$ at  a point $\bttheta$,
\begin{align}
\begin{split}
L(\btheta) &\approx L(\bttheta)+(\btheta-\bttheta)'L'(\bttheta)+
\frac{1}{2}(\btheta-\bttheta)'L''(\bttheta)(\btheta-\bttheta) \\
& \doteq  {\bb(\bttheta) }^\prime \btheta-\half \btheta'\bC(\bttheta)  \btheta
\label{eq:iter}
\end{split}
\end{align}
where $\doteq$ means equality up to an additive
 constant that does not depend on $\btheta$,  and
$L'(\btheta)$ and $L''(\btheta)$ are the first and second partial derivatives of $L(\btheta)$. In equation~\eqref{eq:iter},
$\bb(\btheta) = L'(\btheta)-L''(\btheta)\btheta$ and $\bC(\btheta) =-L''(\btheta)$. The Newton-Raphson
iteration is initialized at the current value of $\btheta$, which we call $\btheta^0 $, and at the $i$th iterate ($i<k$) we expand
$L(\btheta)$ about $\bttheta^{\{i\}} = \bC(\bttheta^{\{i-1\}})^{-1} \bb(\bttheta^{\{i-1\}} )$.
Let $\bttheta^{\{k\}}$ be the ${\{k\}}$th (last) iterate of $\btheta$.
Then the proposal density is the multivariate normal density
$q(\btheta|\btheta^0) = \MVN(\btheta; \bC(\bttheta^{\{k\}})^{-1} \bb (\bttheta^{\{k\}}), \bC(\bttheta^{\{k\}})^{-1})$.
We generate  a candidate point $\btheta^1$ and accept it with probability
\begin{align}\label{eq: accept}
\alpha(\btheta^c\rightarrow\btheta^p)=\min{\left\{1,\frac{\pi(\btheta^1)p(\bz|\btheta^1)
q(\btheta^0|\btheta^1)}
{\pi(\btheta^0)p(\bz|\btheta^0)q(\btheta^1|\btheta^0)}\right\}};
\end{align}
otherwise we take $\btheta^0$ as the new value of $\btheta$.

At equation~ \eqref{eq: accept}, the density $q(\btheta|\btheta^1)$ is constructed using
$k$ iterates starting at $\btheta^1$ in exactly the same way that $q(\btheta|\btheta^0)$
is constructed starting at $\btheta^0$. We note that instead of a
multivariate normal proposal it is straightforward to use a multivariate-t  proposal. In low dimensions,
one or two iterations are sufficient to obtain
good estimation results with reasonable staring values. Extremely poor initial values  will make the \kstep{} proposal
distribution less appropriate, but the poor initial values will also make it much more difficult for the `optimization' method
to converge. We also note that if the \kstep{} method is continued till
convergence, then it is equivalent to the
\lq optimization\rq{}  method and
$q(\btheta|\btheta^0) = q(\btheta|\btheta^1)=\MVN(\btheta; \thetahat ,\Sigmahat)$,
where $\thetahat$ is the mode of $L(\btheta)$ and $\Sigmahat$ is the inverse of the
negative of the Hessian at the mode.

\subsubsection{Delayed rejection using \kstep{} and adaptive MCMC sampling}
We refine the \kstep{} method by using it as a first stage in a \dr{} scheme.
If the proposed value of $\btheta$ generated by the
\kstep{} method is rejected in the first stage then a second value of $\btheta$ is
generated from an \arwm{} proposal in the second stage. The motivation for using the
second stage is to keep the parameters moving by small increments even when the first stage
proposal is poor in certain regions of the parameter space.
The acceptance probability in the second stage takes into account the first
stage rejection to ensure that the posterior distribution is the invariant
distribution of the chain. Our article uses \arwm{} proposal distribution
suggested in \cite{roberts:rosenthal:2009}, although there are a number of
other \arwm{} proposals in the literature, e.g.~\cite{haario:2005}.
Let $\btheta^0$ be the current value of $\btheta$ before \dr{}, $q_1$
the \kstep{} proposal, $\btheta^1$ the value of $\btheta$ proposed by $q_1$
and $\alpha_1$ the first stage acceptance probability at equation~\eqref{eq: accept}.
Let $q_2$ be the \arwm{} proposal density, $\btheta^2$ the value of $\btheta$
proposed by $q_2$ if $\btheta^1$  is rejected and let $\alpha_2$ be the
second stage acceptance probability. Then,
\begin{align*}
q_2(\btheta_0 \rightarrow \btheta) =(1-\delta)\MVN(\btheta;\btheta_0, 2.38^2\bSigma/d)+\delta\hspace{0.1cm}\MVN(\btheta; \btheta_0, 0.1^2\mathbf{I}_d/d)
\end{align*}
and
\begin{align*}
\alpha_2(\btheta^0\rightarrow\btheta^1\rightarrow\btheta^2)& =\min{\left\{1,\frac{\pi(\btheta^2)p(\bz|\btheta^2)q_1(\btheta^1|\btheta^2)
(1-\alpha_1(\btheta^2\rightarrow\btheta^1))}
{\pi(\btheta^0)p(\bz|\btheta^0)q_1(\btheta^1|\btheta^0)
(1-\alpha_1(\btheta^0\rightarrow\btheta^1))}\right\}}\ ,
\end{align*}
where $\bSigma$ is the sample covariance matrix of $\btheta$ estimated from the MCMC iterates obtained thus far,
$\bI_d$ is an identity matrix of dimension $d$, and $\delta$ is a scaling factor which we choose as 0.05. Note that $q_2(\cdot)$ does not enter the second stage acceptance probability because the proposal is symmetric, i.e. $q_2(\btheta_2\rightarrow\btheta_0)=q_2(\btheta_0\rightarrow\btheta_2)$.

 \cite{mira:1999, mira:2001} proposed the \dr{} method
to reduce the number of rejections in \MCMC{}  when a \MH{} proposal is used. In principle the \dr{} method can have more than two stages. In practice, the computational load increases as more stages are added and the acceptance ratios in the later stages are more complex.

\subsubsection{Monte Carlo EM method}\label{sss: univariate mcem}
Most approaches in the literature estimate the \usv{} model using MCMC.
This section  considers a  \mcem{} approach to estimate the parameters
$\btheta = (\mu, \phi, \sigma)^\prime$,  using the centered parameterization of
the \usv{} model at equation~\eqref{eq: uni trans}.

The EM algorithm is described by \cite{mclachlan:krishnan:2008}
and consists of repeated application of expectation (E) and maximization (M) steps.
For our problem, the E-step evaluates
\begin{align} \label{eq:uni-Estep1}
Q(\btheta, \bthetaold) & = \int \log p(\by,\bh|\btheta)p(\bh|\by,\bthetaold)d \bh\ ,
\end{align}
where $\bthetaold$ is the current value of $\theta$.
However, this integral is intractable because the density $p(\bh|\by,\bthetaold)$
is non-Gaussian.

To facilitate the computation, we approximate
$\log(e^2_t)$ as a five component mixture of normals as at equation
\eqref{eq: norm approx} and reexpress equation~\eqref{eq:uni-Estep1} as
\begin{align} \label{eq:uni-Estep2}
Q(\btheta, \bthetaold) & = \int \log p(\by,\bh,\bK|\btheta)p(\bh,\bK|\by,\bthetaold)d (\bh,\bK)\ .
\end{align}
It is difficult to evaluate this integral analytically so we approximate it using the \mcem{}
method, discussed in \cite{wei:tanner:1990}, as
\begin{align} \label{eq:uni-Estep3}
Q(\btheta, \bthetaold) & \approx \frac{1}{M}\sum^{M}_{j=1}\log p(\by,\bh^j,\bK^j|\btheta)
\end{align}
where $(\bh^j,\bK^j), j=1, \dots, M$ are iterates from the joint conditional distribution
$p(\bh,\bK|\by,\bthetaold)$.
Although sampling directly from  $p(\bh,\bK|\by,\bthetaold)$
is difficult, it is straightforward to sample from $p(\bK|\by,\bh,\bthetaold)$ and
from $p(\bh|\by,\bK,\bthetaold)$, so that we use a Gibbs sampling scheme to
obtain a  sample from the joint distribution $p(\bh,\bK|\by,\bthetaold)$
after some burn-in iterations.

Since $\log p(\by,\bh,\bK|\btheta)=\log p(\by|\bh,\bK)+ \log p(\bK)+ \log p(\bh|\btheta)$, and $\btheta$ only enters the last term, the right side of
equation~\eqref{eq:uni-Estep3} is estimated  as
\begin{align*}
Q(\btheta, \btheta^\mathrm{old})&\doteq -\frac{1}{2M}\sum^{M}_{j=1}\sum^{T}_{t=2}\left(\log(\sigma^2)+
\frac{((h^j_t-\mu)-\phi(h^j_{t-1}-\mu))^2}{\sigma^2}\right)\ ,
\end{align*}
where $\doteq$ means excluding additive constants that do not depend on $\theta$. To maximize $Q(\btheta, \bthetaold)$ with respect to
 $ \mu$, $\phi$ and $\sigma^2$, we first reparameterize the centred \usv{}
  as $h_t=\delta+\phi h_{t-1}+\sigma \eta_t$ with $\delta=\mu(1-\phi)$, and then treat the maximization problem as an OLS regression problem for the parameters at the
  $j$-th iterate. The updated parameters in the M-step are
\begin{align}
\begin{split}
\phi^{\mathrm{new}}&=\frac{\sum^{M}_{j=1}\sum^{T}_{t=2}(h^j_t-\overline{h}_t)(h^j_{t-1}-\overline{h}_{t-1})}{\sum^{M}_{j=1}\sum^{T}_{t=2}(h^j_{t-1}-\overline{h}_{t-1})^2}\\
\delta^{\mathrm{new}}&=\overline{h}_t-\phi^{\mathrm{new}}\overline{h}_{t-1}\\
(\sigma^2)^{\mathrm{new}}&=\frac{1}{M(T-1)}\sum^{M}_{j=1}\sum^{T}_{t=2}(h^j_t-\delta^{\mathrm{new}}-\phi^{\mathrm{new}}h^j_{t-1})^2\\
\mu^{\mathrm{new}}&=\delta^{\mathrm{new}}/(1-\phi^{\mathrm{new}})
\end{split}
\label{eq:uni-Mstep}
\end{align}
where
\begin{align*}
\overline{h}_{t}& =\frac{1}{M(T-1)}\sum^{M}_{j=1}\sum^{T}_{t=2}h^j_t \quad \text{and} \quad \overline{h}_{t-1}  =\frac{1}{M(T-1)}\sum^{M}_{j=1}\sum^{T}_{t=2}h^j_{t-1} \ .
\end{align*}

A difficulty in using the \mcem{} method is monitoring convergence,
since approximation errors entering in the E-step  mean that the usual ways  of determining convergence such as stopping when the change in the parameters is small or the change in the log-likelihood is small may be unreliable. We propose to monitor convergence
by using the bridge sampling approach of \cite{meng:schilling:1996} to estimate the likelihood ratio $p(\by|\btheta^{i})/ p(\by|\btheta^{i-1})$  at the $i$th iteration of the \mcem{} by
\begin{align}
R(\btheta^{i},\btheta^{i-1})=\frac{\sum^{M}_{j=1}\biggl (p(\by,\bh^{\{i-1,j\}},
\bK^{\{i-1,j\}}|\btheta^{i})/
p(\by,\bh^{\{i-1,j\}},\bK^{\{i-1,j\}}|\btheta^{i-1})\biggr )^{\half}}
{\sum^{M}_{j=1}\biggl (p(\by,\bh^{\{i,j\}},
\bK^{\{i,j\}}|\btheta^{i-1})/
p(\by,\bh^{\{i,j\}},\bK^{\{i,j\}}|\btheta^{i})\biggr )^{\half}}
\label{eq:uni-lr}
\end{align}
where $(\bh^{\{i,j\}},\bK^{\{i,j\}}, j=1, \dots , M) $ are  generated from
$p(\bh,\bK|\by, \btheta^{i })$ at the $i$th \mcem{} iteration.
Following \cite{meng:schilling:1996} we say that \mcem{} iteration has converged if the plot of $R(\btheta^{i},\btheta^{i-1})$ against $i$ converges to 1 from above.

We obtain approximate large sample standard errors for the parameters using a
method by Louis \citep[see][p. 226]{mclachlan:krishnan:2008}.
The approximate information matrix of the observed log-likelihood is
\begin{align}
\begin{split}
\bI(\btheta;\by)&\approx -\frac{1}{M}\sum^{M}_{j=1}\frac{\partial^2\log p(\by,\bh^j,\bK^j|\btheta)}{\partial\btheta\partial\btheta'}\\
&-\frac{1}{M}\sum^{M}_{j=1}\left(\frac{\partial\log p(\by,\bh^j,\bK^j|\btheta)}{\partial\btheta}\right)\left(\frac{\partial\log p(\by,\bh^j,\bK^j|\btheta)}{\partial\btheta'}\right)\\
&+\left(\frac{1}{M}\sum^{M}_{j=1}\frac{\partial\log p(\by,\bh^j,\bK^j|\btheta)}{\partial\btheta}\right) \left(\frac{1}{M}\sum^{M}_{j=1}\frac{\partial\log p(\by,\bh^j,\bK^j|\btheta)}{\partial\btheta'}\right)
\label{eq:uni-informatrix}
\end{split}
\end{align}

We note that \cite{shephard:1993} uses a similar Monte Carlo EM approach
to estimate a centered \usv{} model. The major differences between our treatment and
that of \cite{shephard:1993} are (i) our approach for estimating the right side
of equation~\eqref{eq:uni-Estep2} is  more effective than the single-site Gibbs sampling
and Muller's random walk proposal to sample from $p(\bh|\by,\btheta)$  proposed in
\cite{shephard:1993}. The extra efficiency of our approach is  especially important in
\fmsv{} models.   (ii) \cite{shephard:1993}   monitors convergence using a traditional EM
approach instead of using the likelihood ratios obtained from the bridge  sampling method.
(iii)~Our main focus is on the feasibility of the \mcem{} approach for
\fmsv{} models discussed in section~\ref{Sec: factor msv models}, whereas \cite{shephard:1993} only considers the univariate case.

\vspace{0.5cm}
\section{Computational methods for \fmsv{} models}\label{Sec: factor msv models}
There are a number of \fmsv{} models in the literature.
To illustrate the estimation methods in our paper, we consider the following \fmsv{} model,
\begin{align}
\begin{split}
\by_t & =\bB \bmf_t+\bS^{1/2}_t\be_t, \hspace{0.2cm}
\bS_t=\textrm{diag}\left\{\exp(h_{1,t}),...,\exp(h_{p,t})\right\}, \hspace{1cm} \be_t\sim
N(0,\bI_p)\\
\bmf_t & =\bV^{1/2}_t\bu_t, \hspace{1.2cm}
\bV_t=\textrm{diag}\left\{\exp(h_{p+1,t}),...,\exp(h_{p+k,t})\right\},
\hspace{0.2cm} u_t\sim N(0,\bI_k)\\
h_{j,t} & =\mu_j+\phi_j(h_{j,t-1}-\mu_j)+\sigma_j\eta_{j,t},  \hspace{4.6cm}
\eta_{j,t}\sim N(0,1) \label{eq:multi}
\end{split}
\end{align}
where both the latent factors and the idiosyncratic shocks are allowed to follow
different stochastic volatility processes, $\bB$ is the factor loading matrix,
$p$ and $k$ are the number of return series and factors, with
$\by_t=(y_{1,t},...,y_{p,t})'$ and $\bmf_t=(f_{1,t},...,f_{k,t})'$.
To identify the model we follow \cite{chib:nardari:shephard:2006}
 and set $b_{ii}=1$ for $i\leq k$ and $b_{ij}=0$ for $j>i$.
We also assume both the idiosyncratic shocks and the latent factors are
internally and mutually uncorrelated so that $\bS_t$ and $\bV_t$ are diagonal
matrices. Consequently, the correlation structure of
$\by_t$ is governed by the latent factors $\bmf_t$.

We first review the sampling scheme for estimating the \fmsv{} model parameters
in \cite{chib:nardari:shephard:2006}, which is also used by \cite{han:2006}
and \cite{nardari:scruggs:2006} in their applications.  We again refer to it
as the `optimization' MCMC method because optimization is executed within each of the
MCMC iterations similarly to its use in the \usv{} model.

\subsection{Optimization MCMC Sampling Method}
\begin{enumerate}
\item Initialize $\btheta=(\btheta_{1},...,\btheta_{p+k})$ with $\btheta_{j}=(\phi_j,\sigma_j)'$,  $\bh=(\bh_1,...,\bh_{p+k})'$ with each $\bh_j$ defined as in \usv{} case.
\item Jointly sample $\bbeta,\bmf\sim p(\bbeta,\bmf|y,\bh)$, where $\bbeta$ represents the free parameters in the $\bB$ matrix.
		\begin{enumerate}[2.1.]
		\item Sample $\bbeta\sim p(\bbeta|\by,\bh)$
				\begin{enumerate}[a.]
				\item Sample the candidate value $\bbeta^1\sim
q_T(\betahat|{\Sigmahat}_\beta,v)$, where $\betahat$ is the mode
maximizing the log-likelihood function $\log p(y|\bbeta,\bh)$, $\Sigmahat_\beta$ is the corresponding covariance matrix at the mode and $v$ is degree of freedom for the multivariate-t distribution.
				\item Accept the candidate value $\bbeta^1$ with probability
				\begin{align*}
\alpha(\bbeta^0\rightarrow\bbeta^1)=\min{\left\{1,\frac{\pi(\bbeta^1)
p(\by|\bbeta^1)q_T(\bbeta^0|\widehat{\bbeta},\Sigmahat_\beta,v)}
{\pi(\bbeta^0)p(\by|\bbeta^0)q_T(\bbeta^1|\betahat,\Sigmahat_\beta,v)}\right\}}
				\end{align*}		
				\end{enumerate}	
		\item Sample the latent space-state vector  $f\sim p(\bmf|\by,\bB,\bh)$ using a Forwards-Filtering-Backwards-Sampling (FFBS) algorithm.
		\end{enumerate}
\item Sample the indicator vector variables $\bK\sim p(\bK|\by,\bB,\bmf,\bh)$, noting that
given $(\bB,\bmf)$, we can decompose the \fmsv{} model into the $(p+k)$ \usv{} models
				\begin{center}
				$z_{j,t}=\left\{\begin{array}{c} \hspace{-7mm} \log(y_{j,t}-\bB_{j}f_t)^2=h_{j,t}+\log(e^2_{j,t}), \hspace{0.2cm} j\leq p\\
				\log(f_{j-p,t})=h_{j-p,t}+\log(u^2_{j-p,t}), \hspace{0.4cm} j\geq p+1 \end{array}\right.$
				\end{center}
where $\bB_j$ is the $j$-th row of the factor loading matrix $\bB$. The indicator vectors
 $\bK_j$ are now sampled independently through $\bK_j\sim p(\bK_j|\bz_j,\bh_j)$ for $j=1,...p+k$ as for the \usv{} model.
\item Jointly sample $\btheta,\bh \sim p(\btheta,\bh|\by,\bB,\bmf,\bK)$ from $p+k$ separate series as $\btheta_j,\bh_j\sim p(\btheta_j,\bh_j|\bz_j,\bK_j)$  for $j=1,...,p+k$, each of which can be dealt with as in the \usv{} case. That is we sample $\btheta_j\sim p(\btheta_j|\bz_j,\bK_j)$ using \optim{} MCMC and then sample $\bh_j\sim p(\bh_j|\bz_j,\btheta_j,\bK_j)$.
\item Go to Step 2.
\end{enumerate}

The main advantage of the `optimization' method is that it jointly samples $\bbeta,\bmf\sim p(\bbeta,\bmf|\by,\bh)$ and jointly samples $\btheta_j,\bh_j\sim p(\btheta_j,\bh_j|\bz_j,\bK_j)$ for $j=1,...,p+k$. The general Gibbs-sampling method which samples $\bbeta$ conditional on $\bmf$ and then sample $\bmf$ conditional on $\bbeta$ is less efficient as demonstrated in various simulation examples discussed in \cite{chib:nardari:shephard:2006}. The \optim{} method is computationally slow, especially when $\by$ and $\bF$ are high  dimensional,  because now optimization is  used for both $\bbeta$ (which is usually high dimensional) and also for   $\btheta=(\btheta_1,...,\btheta_{p+k})^\prime$ for each \usv{} parameter space.

\subsection{\kstep{}  and delayed rejection method}
We extend the \dr{} method for sampling parameters in the \usv{} model to the \fmsv{} model
using the same basic ideas. Instead of using numerical optimization to find the proposal distributions
based on the  mode, we use the \dr{} method discussed in
section~2 to build the proposal distributions for
$\bbeta$ and the $\btheta_j$ for $j=1,...,p+k$. That is, we first use the \kstep{}  method
to build the proposal after one or two iterations in the first stage, and then use the \arwm{} method in the second stage if the candidate value in the first stage is rejected.
 The computational load in the \fmsv{} model is mainly from sampling $\bbeta\sim p(\bbeta|\by,\bh)$.
 The terms  $\bK_j,\btheta_j,\bh_j$,  for $j=1,...,p+k,$ are sampled independently of each other as discussed above and, if sufficient computing resources are available, they can be sampled in parallel. If $\beta$ is high dimensional, say 50-dimensional,
using the \optim{} method  to find the mode of $\log p(\by|\bbeta,\bh)$ is
slow . Although we can easily build the proposal distribution through the \kstep{} method by only running
 one or two Newton-Raphson iterations, we may encounter  high
 rejection rates in this stage.  Therefore,
 the efficiency of the \MCMC{} relies more on the \arwm{} in the second stage
 which does not work as well in high dimensions.
 To reduce problems in high dimensions, we make two suggestions.
 First, increase the number of iterations
 in the first stage, e.g. to five or six iterations,
 but in that case the speed advantage of the \dr{} method is reduced.
 Second, split $\bbeta$ into several smaller sub-blocks such as $\bbeta=(\bbeta'_1,...,\bbeta'_r)^\prime$, with each sub-block
 containing a relatively small number of parameters.  For example, for a 50-dimensional $\bbeta$,
 we can use a total of 7 sub-blocks with 8 parameters for the first 6 sub-blocks and 2 parameters for the
 last sub-block and update $\bbeta$ using the scheme
$\bbeta^{(j)}_i \sim p(\bbeta_i|\by,\bh,\bbeta_l^{(j)}, l < i, \bbeta^{(j-1)}_l, l>j)$ for $i=1, \dots, r$.
where $j$ represents the current iteration.
By sampling a large dimensional $\bbeta$ in blocks, the performance of both the `optimization' and \dr{}  methods can
improve,  especially the \dr{} method, where one or two iterations in the first stage may be sufficient to obtain good results.

\subsection{Monte Carlo EM Method for \fmsv}\label{ss: multivariate mcem}
 We now show how the \mcem{} method discussed in section~\ref{sss: univariate mcem}
 for the \usv{}  model generalizes in a straightforward manner to the \fmsv{}. The E-step evaluates the complete log likelihood as
\begin{align}
Q(\bTheta, \bTheta^\mathrm{old}) & = \int \log p(\by,\bF,\bh,\bK|\bTheta)p(\bF,\bh,\bK|\by,\bTheta^\mathrm{old})d (\bF,\bh,\bK)\ ,
\label{eq: msv-Estep1}
\end{align}
where $\bTheta=\left\{\bbeta, \btheta\right\}$, $\by=(\by_1,...,\by_p)$, $\bF=(\bmf_1,...,\bmf_k)$, $\bh=(\bh_1,...,\bh_{p+k})$ and $\bK=(\bK_1,...,\bK_{p+k})$ represent matrices with each column series starts from $t=1$ to $T$. Similarly
 to the \usv{}, we approximate the integral using Monte Carlo simulations as
\begin{align}
Q(\bTheta, \bTheta^\mathrm{old}) & \approx \frac{1}{M}\sum^{M}_{j=1}\log p(\by,\bF^j, \bh^j,\bK^j|\bTheta)
\label{eq: E-step6}
\end{align}
where $\log p(\by,\bF,\bh,\bK|\bTheta)=\log p(\by|\bF,\bh,\beta)+\log p(\bF|\bh)+\log p(\bK)+\log p(\bh|\btheta)$ with $\beta$ and $\btheta$ entering the first and last terms. The $(\bF^j, \bh^j,\bK^j)$, for $j=1, \dots, M$,
are generated
 from the joint posterior $p(\bF,\bh,\bK|\by,\bTheta^\mathrm{old})$ and are obtained using Gibbs sampling
 by first generating from $p(\bK|\by,\bF, \bh)$, then from $p(\bh|\by,\bF,\bK, \btheta)$ and finally from $p(\bF|\by,\bh,\beta)$ after some burn-in iterates.

Because $\bbeta$ only enters $\log p(\by|\bF,\bh,\bbeta)$, maximizing the expected complete log likelihood with respect with $\bbeta$ is equivalent to maximizing
\begin{align*}
\begin{split}
Q(\bTheta, \bTheta^\mathrm{old}) &\doteq \frac{1}{M}\sum^{M}_{j=1}\log p(\by|\bF^j, \bh^j, \bbeta)\\
&=\frac{1}{M}\sum^{M}_{j=1}\sum^{p}_{i=1} \log p(\by_i|\bF^j, \bh^j_i, \bB_i)\\
&=\frac{1}{M}\sum^{M}_{j=1}\sum^{p}_{i=1}\sum^{T}_{t=1}-\half\left(h^j_{i,t}+\frac{(y_{i,t}-\bB_i \bF^j_t)^2}{\exp(h^j_{i,t})}\right)\ .
\end{split}
\end{align*}
The second line of the equation follows from the first because the $\by_i$ series are independent conditional on
$\bF$ and $\bh$. The rows of the optimal $\bbeta$ are
\begin{align*}
\bbeta^{\mathrm{new}}_i=\left(\frac{1}{M}\sum^{M}_{j=1}\sum^T_{t=1}\bF^{*j}_t\bF^{*j'}_t\right)^{-1}\left(\frac{1}{M}\sum^{M}_{j=1}\sum^T_{t=1}y^{*}_{i,t}\bF^{*j'}_t\right)
\hspace{0.5cm} \mathrm{for}\hspace{0.1cm} i=2,...,p,
\end{align*}
where $y^{*}_{i,t}=y_{i,t}$, $\bF^{*j}_{t}=\bF^{j}_t$ for $i>k$, and $y^{*}_{i,t}=y_{i,t}-B_{i,i: k}\bF^j_{i:  k,t}$, $\bF^{*j}_{t}=\bF^{j}_{1: i-1, t}$ for $2\leq i \leq k$. With $\bF_t=(f_{1,t},...,f_{k,t})'$, $B_{i, i: k}=(B_{i, i},...,B_{i, k})$, $\bF_{i:  k, t}=(f_{i,t},...,f_{k,t})'$ and $\bF_{1:  i-1, t}=(f_{1,t},...,f_{i-1,t})'$.

 Because $\btheta$ only enters $\log p(\bh|\btheta)$, maximizing the expected complete log likelihood respect with $\btheta$ is equivalent to maximizing
\begin{align*}
\begin{split}
Q(\bTheta, \bTheta^\mathrm{old}) &\doteq \frac{1}{M}\sum^{M}_{j=1}\log p(\bh^j|\btheta)=\frac{1}{M}\sum^{M}_{j=1}\sum^{p+k}_{i=1} \log p(\bh^j_i|\btheta_i)\\
&=-\frac{1}{2M}\sum^{M}_{j=1}\sum^{p+k}_{i=1}\sum^T_{t=2}\left(\log(\sigma^2_i)+\frac{((h^j_{i,t}-\mu_i)-\phi_i(h^j_{i,t-1}-\mu_i))^2}{\sigma^2_i}\right)
\end{split}
\end{align*}
Since each of the series $\bh_i, i=1,...,p+k$ is conditionally independent,
each $\btheta^{\mathrm{new}}_i=(\mu^{\mathrm{new}}_i,\phi^{\mathrm{new}}_i,\sigma^{\mathrm{new}}_i)$ is  obtained  as in the \usv{} case at equation~(\ref{eq:uni-Mstep}).

Convergence is  monitored in the \usv{} case as at equation~(\ref{eq:uni-lr}),
except that we now have the complete log likelihood as $\log p(\by,\bF,\bh,\bK|\bTheta)$. The information matrix can be calculated using a similar formula to equation (\ref{eq:uni-informatrix}) by substituting in the current complete log likelihood.

\section{Marginal Likelihood calculation} \label{s: marginal likelihood}
 An important practical issue in estimating factor MSV models is determining the
  number of latent factors. A Bayesian approach to this problem is to choose the number of factors using marginal likelihood.  Usually there is no closed form solution available for the marginal likelihood in the \fmsv{} model and it is necessary to estimate the marginal likelihood using simulation methods such as \cite{chib:1995},
  where additional `Reduced MCMC' runs are needed. For a given \fmsv{} model, an expression for the
  marginal likelihood is obtained through the identity
\begin{align}\label{eq: ML fmsv}
p(\by) & =\frac{p(\by|\btheta^*,\bbeta^*)p(\btheta^*,\bbeta^*)}{p(\btheta^*,\bbeta^*|\by)},
\end{align}
where $(\btheta^*,\bbeta^*)$
are chosen as high density ordinates such as posterior means or componentwise posterior medians, with $\btheta^*=(\btheta^*_1,...,\btheta^*_{p+k})$, $\btheta^*_j=(\mu^*_j,\phi^*_j,\sigma^*_j)'$ for $j=1,...,p+k$.
Our article  uses componentwise posterior medians. Such a choice also has advantages for the copula methods that are discussed later. In particular, we choose $\bbeta^*$ as the componentwise posterior median from the MCMC simulation output in the estimation stage and $\btheta^*$ as the componentwise posterior median from later `Reduced MCMC' runs. The likelihood $p(\by|\btheta^*,\bbeta^*)$ in the numerator of
equation~\eqref{eq: ML fmsv} can be evaluated sequentially by integrating out the latent $\bh_t$ using the auxiliary particle filter introduced by \cite{pitt:shephard:1999} and
illustrated in detail in \cite{chib:nardari:shephard:2006}.
The density $p(\btheta^*,\bbeta^*)$ in equation \eqref{eq: ML fmsv} is the prior density. The denominator $p(\btheta^*,\bbeta^*|\by)$ in equation \eqref{eq: ML fmsv} is  decomposed as
\begin{align*}
p(\btheta^*,\bbeta^*|\by)=p(\bbeta^*|\by)p(\btheta^*|\by,\bbeta^*)\ ,
\nonumber
\end{align*}
where the posterior marginal density $p(\bbeta^*|\by)$ is approximated by a normal distribution with mean and covariance matrix obtained from the MCMC runs in the estimation stage. \citep[][also use the same approximation]{chib:nardari:shephard:2006} . Alternatively, one can use the Gaussian copula method discussed later to estimate the posterior ordinate. To estimate the second term $p(\btheta^*|\by,\bbeta^*)$, it is necessary to run at least one \lq Reduced MCMC\rq{} to obtain a sampler of $\btheta$ from $p(\btheta|\by,\bbeta^*)$. Generally, this posterior conditional density does not have a closed form and it is necessary to use nonparametric methods such as kernel density estimation \citep[e.g.][]{terrell:1990} to evaluate the ordinate. However, the dimension of $\btheta^*$ is usually large, and  kernel density estimates may not be accurate enough for such large dimensional cases. One way to obtain a reasonably good estimate is to split $\btheta^*$ into several smaller sub-blocks as in \cite{chib:nardari:shephard:2006}
 where two $\btheta^{*}_j$'s are put in one sub-block as in
\begin{align}\label{eq: beta blocks}
p(\btheta^*|\by,\bbeta^*)=p(\btheta^*_1,\btheta^*_2|\by,\bbeta^*)p(\btheta^*_3,
\btheta^*_4|y,\bbeta^*,\btheta^*_1,\btheta^*_2)\times\cdots\times p(\btheta^*_{p+k-1},\btheta^*_{p+k}|\by,\bbeta^*,\btheta^*_1,...,\btheta^*_{p+k-2})
\end{align}
By splitting $p(\btheta^*|\by,\bbeta^*)$ into blocks as in
equation~\eqref{eq: beta blocks},
$\textrm{Max}((p+k)/2)$ \lq Reduced MCMC\rq{}
 runs are necessary, where $\textrm{Max}$ is the smallest integer greater
 than or equal to $(p+k)/2$; we note that
 additional sub-blocks of parameters are  fixed at  successive \lq Reduced MCMC\rq{}
  runs as in \cite{chib:1995}.
The computational time necessary to carry out this sequence of
\lq Reduced MCMC\rq{}  runs is large and usually much greater
than the time required for estimating the model.

Our article adopts the copula-based approximation method proposed by \cite{nott:kohn:xu:fielding:2009}
to evaluate posterior ordinates. This method  can deal with  larger dimensional blocks of
parameters than  kernel density estimation methods.
The Gaussian copula   approximation to  a multivariate density $f(\bzeta)$, with $\bzeta$ a $p \times 1 $ vector, is
\begin{align} \label{eq: gaussian copula}
q(\bzeta)=|\bC|^{-1/2}\exp\left(\frac{1}{2}\bEta'\left(\bI(p)-\bC^{-1}\right)\bEta\right)
\prod_{j=1}^{p}f_j(\zeta_j)
\end{align}
where $\bC$ is the correlation matrix of the Gaussian copula, $\bEta=(\eta_1,...,\eta_p)'$ with $\eta_j=\Phi^{-1}(F_j(\zeta_j))$, $\Phi$ is  the  standard normal CDF, $F_j$ is the CDF of the $j$th marginal $\zeta_j$ and $f_j$ is the corresponding density.
When $\zeta_j$ is the posterior median of the $j$ marginal, then $\eta_j = 0 $ and equation
\eqref{eq: gaussian copula} simplifies to
\begin{align} \label{eq: gaussian copula median}
q(\zeta)=|C|^{-1/2}\prod_{j=1}^{p}f_j(\zeta_j)\ .
\end{align}

Suppose that we wish to approximate $p(\btheta^*|\by,\bbeta^*)$ using one block and that  $\btheta^*$ is the componentwise posterior median of the $\btheta$ iterates obtained from $p(\btheta|y,\bbeta^*)$. Then, from equation~\eqref{eq: gaussian copula median}  the estimate of the posterior ordinate is
\begin{align*}
p(\btheta^*|\by,\bbeta^*)\approx|C|^{-1/2}\prod_{j=1}^{p+k}
\prod_{i=1}^{3}f_{j,i}(\btheta^*_{j,i}|\by,\bbeta^*)
\end{align*}
where $f_{j,i}(\btheta^*_{j,i}|\by,\bbeta^*)$ with $\btheta^*_j=(\mu^*_j,\phi^*_j,\sigma^*_j)'$ represents the one dimensional marginal density ordinate which can be estimated using kernel density estimation. The copula correlation matrix $\bC$ is estimated
using order statistics as in \cite{nott:kohn:xu:fielding:2009}.

\section{Simulation Study}\label{s: simulation study}
This section provides several simulation examples for both \usv{} and \fmsv{}
models that illustrate  the computational methods discussed in
sections~\ref{sec:unimodel} and \ref{Sec: factor msv models}.
The methods are compared using performance diagnostics.

\subsection{Performance diagnostics}\label{ss: performance diagnostics}
A popular way to inspect the performance of the MCMC samplers is based on the  inefficiency factors for the parameters. For any given parameter $\theta$, the inefficiency factor is defined as
\begin{align}\label{eq: ineff}
\textrm{Inefficiency factor} & =1+2\sum_{j=1}^{G-1}\left(1-\frac{j}{G}\right)\rho(j)\ ,
\end{align}
where $G$ is the number of sample iterates of $\theta$ and $\rho(j)$ is
the $j$-th  autocorrelation of the iterates of $\theta$. When $G$ is large
and $\rho(j)$ tends to zero quickly, the inefficiency factor is approximated by $1+2\sum_{j=1}^J\rhohat (j)$, for some given $J$, with $\rhohat (j)$ the $j$th sample autocorrelation  of $\theta$.
We choose $J=100$ in all the empirical analyses because
$\rhohat (j)$ usually decays to zero before 100 lags for almost all the parameters in our simulation examples.

Mathematically, the inefficiency factor is just the ratio of the variance of a posterior mean of the iterates obtained from MCMC sampling to the variance of the posterior mean from independent sampling. The inefficiency factor is interpreted
as that multiple of the number of iterates $G$ that gives the same accuracy as $G$ independent iterates. A low inefficiency factor
is preferred to a higher inefficiency factor.

The inefficiency factor itself may not be informative enough to compare two sampling methods because it does not take computation time into account. Consider, for example, two sampling methods that have very similar inefficiency factor scores, but the first method is computationally faster than the second one. We then say that the first method is relatively more efficient because it needs less computational time to achieve the same accuracy as the second method. To take into account both the inefficiency factor and the computing time we consider the equivalence factor score defined as
\begin{align} \label{eq: equivalence factor}
\textrm{Equivalence factor}=\textrm{inefficiency factor}\times t,
\end{align}
where $t$ is the computing time per iteration. For a given parameter, the ratio of equivalence factors for two sampling schemes is
the ratio of times taken by the two sampling schemes to obtain the same accuracy for a given parameter.
We note that although the equivalence factor gives a more complete picture of the performance of a sampler, it is implementation dependent.

\subsection{Simulation examples of univariate stochastic volatility models}
\label{ss: usv simulation examples}
The \usv{} model at equation \eqref{eq:uni} is estimated using the \optim{},
\kstep{}, \dr{} and \mcem{} methods discussed in section~\ref{sec:unimodel}
for two simulated datasets (each having 10 replicates) with sample sizes of 500 and
1,500, respectively.
We use $k = 1$ steps in the first stage for both the \kstep{} and the \dr{} methods.
The true parameters for the data sets of 500 and 1,500 observations
are $(\mu=0.5,\phi=0.9,\sigma=0.1)$ and $(\mu=1.0,\phi=0.95,\sigma=0.15)$.
The parameter $\mu$ is nested in latent the state vector $h_t$ as in
equation \eqref{eq:uni2}.  The same prior specifications are used for both data sets
\begin{align*}
& \mu\sim N(0,5), \hspace{0.5cm} \phi\sim \textrm{Beta}(8,0.1), \quad
\text{and} \quad   \sigma\sim \rmIG(2,0.1),
\end{align*}
where the persistence  parameter $\phi$ ranges between 0 and 1 because the volatility
is  positively autocorrelated in most financial time series;  $\rmIG(a,b)$
is  the inverse Gamma distribution with shape parameter $a$ and scale
parameter $b$ (and with mean $b/(a-1))$. All MCMC simulation results are computed using
the last 5,000 draws after discarding the first 1,000 burn-in iterations.
For \mcem{}, we set  $M=100$ after 10 burn-in iterations. Extensive testing using
longer burn-in periods and higher values of $M$  produced similar results.

Table~\ref{tab:table1} summarizes the estimation results showing
that  the three MCMC methods provide accurate parameter estimates for both datasets.
The inefficiency factors of the three \usv{} parameters are all less than 20 for all
MCMC methods across both datasets. The \dr{} method has the lowest inefficiency factors
for  the dataset of 500 observations and the \optim{} method has the
lowest inefficiencies for the dataset with 1500 observations.
However, due to its relatively fast computing speed, the
equivalence factors of the \dr{} method are about half those of the
\optim{} method. In general, the \dr{} method  produces lower inefficiencies
and higher acceptance rates than the \kstep{} method at little additional
computational cost, and it therefore has lower equivalence factor scores than
the \kstep{} method in the most cases.

We run 25 iterations for each replication of the \mcem{} method.
Figure~\ref{fig:fig1} plots the likelihood ratios against the iterate numbers and
 shows that likelihood ratios stabilizes before 25 iterations. We therefore  take the values of the 25-th iterate as the final estimates. Table~\ref{tab:table1} shows that the \mcem{} method performs well in terms of its accuracy  and is promisingly fast in terms of computing speed with only a few minutes of computing time needed compared with hours using the MCMC methods. Therefore, we  expect that by taking the \mcem{} parameter estimates as initial values for the  MCMC methods we will obtain
a chain that converges quickly. The benefits may not be important in the \usv{} model since both the \optim{} and \dr{} methods converge  quickly, but using \mcem{} is worthwhile for the \fmsv{} model.

\subsection{Simulation examples of factor multivariate stochastic volatility models}
\label{ss: simulation fmsv model}
 This section fits the \fmsv{} model at  equation \eqref{eq:multi} to two simulated
 datasets  using 5 replicates for each dataset.
 The first dataset  which we call `P5-K1' has $p$=5 and $k$=1 and the second dataset
 which we call `P10-K2' has $p$=10 and $k$=2. Both datasets have 500 observations.
 The three computational methods (with one-step iteration in the first stage for
 \dr{}) discussed in section~3 are used to estimate the model. The true parameter values
 are $\btheta_j=(0.5,0.9,0.1)'$ for $j=1,...,p$ and $\btheta_j=(1.0,0.95,0.15)'$
 for $j=p+1,...,p+k$. Table~\ref{T: B entries} gives the elements in the loading matrices $\bB$.

The prior specification for both datasets is
$\mu_j\sim N(0,5),   \phi_j\sim \textrm{Beta}(8,0.1), \sigma_j\sim \rmIG(2,0.1)$, for
 $j=1, \dots, p+k$ and $\bbeta\sim N(\mathbf{0},10\textbf{I}_d)
$
with $\textbf{I}_d$ the $d$ dimensional identity matrix.

For model  \lq P5-K1\rq{},
 we use a one block strategy to sample $\bbeta$ from $p(\bbeta|\by,\bh)$ for
both MCMC methods. For the higher dimensional \lq P10-K2\rq{}
model, we use a one block strategy for the \optim{} MCMC method but use a
sub-block strategy for the \dr{} MCMC method, with 8 parameters in each sub-block,
where the last sub-block contains the remaining parameters if $d$ is not a multiple
of  8. All the MCMC simulation estimates  are again computed using the last 5,000 draws
after discarding the first 1,000 burn-in iterations, and $M$ is set as 100 after 10 burn-in iterations for \mcem{}. For the \mcem{} method, similar results were obtained for longer burn-in periods and larger values of $M$.

\begin{figure}[t]
	\centering
		\includegraphics[width=1.0\textwidth]{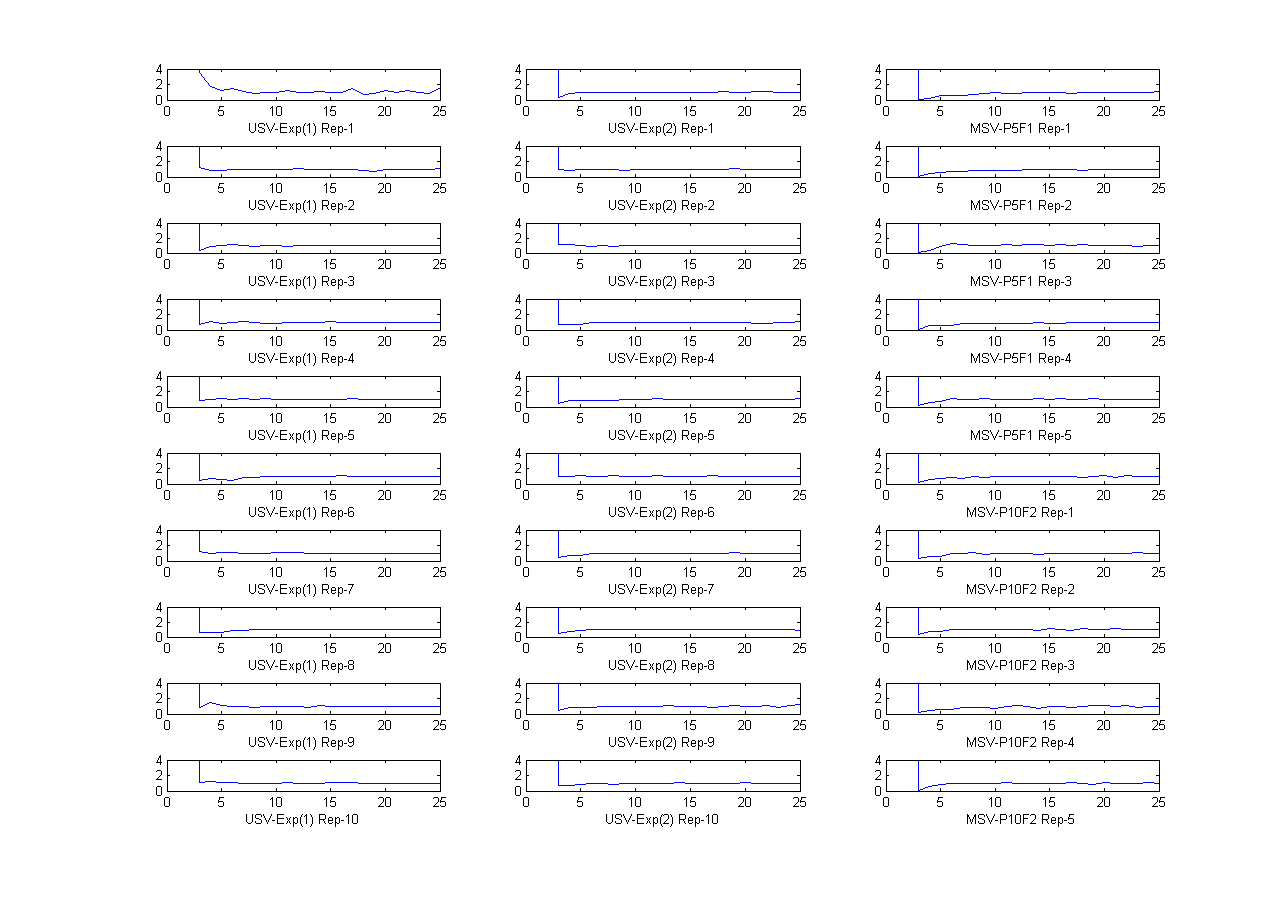}
	\footnotesize
	\caption{{\bf Plots of likelihood ratio against iteration number for \mcem.} The left column of panels
plots the likelihood ratios  for each replicate of the  first \usv{} simulated example (USV-Exp(1)). The middle column
plots the likelihood ratios  for each replicate of the  second  \usv{} simulated example (USV-Exp(2)). The right column plots the likelihood ratios  for each replicate of the \fmsv{} models `P5-F1' example (`MSV-P5F1') and `P10-F2' example (`MSV-P10F2'). The likelihood ratios for the first two iterates are left out because sometimes they
are very large, especially in the \fmsv{} case.}
	\label{fig:fig1}
\end{figure}

Table~\ref{tab:table2} summarizes the estimation results for the \lq P5-K1\rq{}
 model. The estimates from the two MCMC methods agree  for
almost all the unknown parameters and are  also very close to the true parameter values.
The inefficiency factor scores are similar  for $\btheta$ but not for $\bbeta$,
where the inefficiency of the \optim{} method (7.02) is  less than half of the value for
 the  \dr{} method (14.53).  However, due to the faster sampling speed of  the
 \dr{}  method, the \optim{} method has higher equivalence factor scores
 than \dr{}, with the highest value being 48.71 for $\phi$ compared with the
 corresponding value of 20.37  using \dr{} (this is almost 2.4 times higher).
 The parameter estimates from  \mcem{} for $\btheta$
 are consistent with those obtained by MCMC, but a little lower than the MCMC estimates
 for $\bbeta$. Figure~\ref{fig:fig1}  plots the likelihood ratios
 (see equation \eqref{eq:uni-lr}) vs iteration number for the `P5-K1' model for all
 the replicates. The plots suggest that the \mcem{} iterates have converged.
 From Table~\ref{tab:table2},  the \mcem{} method took 6 minutes compared
 with the 5.1 hours for the \optim{} MCMC method.

Table \ref{tab:table3} summarizes the estimation results of the \lq P10-K2\rq{}
 model.  The parameter estimates of the two MCMC methods are again close to the true
parameter values. The inefficiency factors are  similar for both MCMC methods,
but the  \dr{} method has  lower equivalence factor scores (by a factor of 5) than
the  \optim{} method because it is five times faster. The acceptance ratios
for $\theta=(\phi,\sigma)'$ are almost the same for both MCMC methods. We note that the
parameter $\mu$ does not have an acceptance ratio since it is sampled jointly with the latent state $h$ from its exact conditional density. However, we observe a significant decrease for the acceptance ratio $\bbeta$ from 0.87 to 0.66
in the \optim{} method when a one block strategy is used for both simulation cases.
Similar results (that are not reported in the article  but that are available from
the authors) are found for \dr{} method, with a decrease from 0.77 to 0.32 for the
first stage acceptance ratio if a one block strategy is used in both
examples. Tables~\ref{tab:table2} and \ref{tab:table3} show that for the \dr{} method
where  a one-block approach is used in the \lq P5-K1\rq{}
 case but several sub-blocks are used in the \lq P10-K2\rq{}
  case, there is a small
decrease, e.g., from 0.77 to 0.72 in the first stage acceptance ratio,
but the acceptance ratio is even higher than the value of 0.66 for the  \optim{} method
where one block is used in both cases. The benefit of the sub-block strategy in sampling
high dimensional $\bbeta$ is clear in this example.

\begin{figure}[t]
	\centering
		\includegraphics[width=1\textwidth]{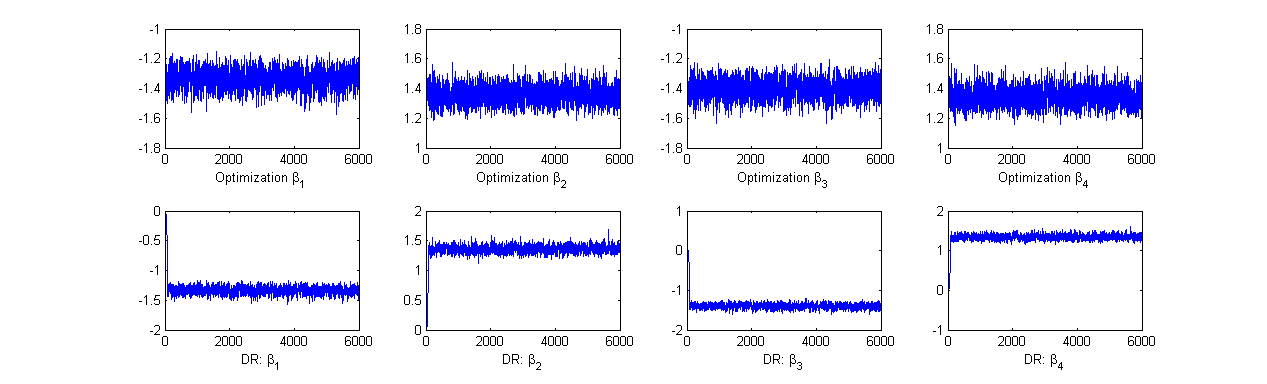}
	\footnotesize
	\caption{{\bf Plots of the iterates for $\beta_1$ to $\beta_4$ for the `P5-F1' simulation example}}
	\label{fig:fig2}
\end{figure}

\begin{figure}[t]
	\centering
		\includegraphics[width=1\textwidth]{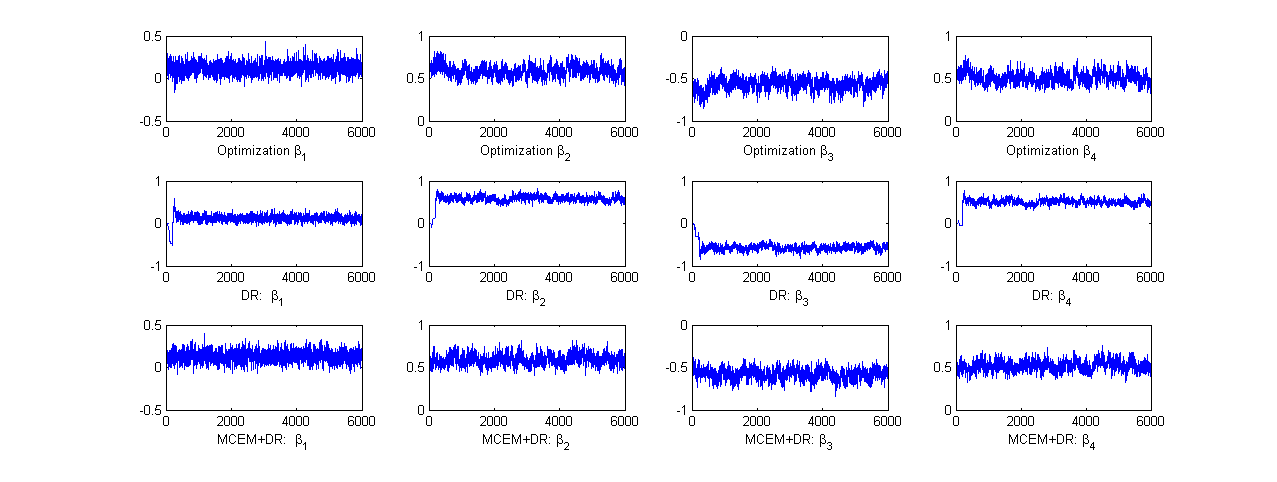}
	\footnotesize
	\caption{{\bf Plots of the iterates for $\beta_1$ to $\beta_4$ for the `P10-F2' simulation example}}
  \label{fig:fig3}
\end{figure}

Table~\ref{tab:table3} also shows that the estimates  from the \mcem{} method are similar to those obtained by both MCMC methods for most of the parameters. Figure~\ref{fig:fig1} plots the likelihood ratios  vs iteration number for the \lq P10-K2\rq{}
 model for all the replicates. The figure suggests that the \mcem{} iterations have converged. Figure~\ref{fig:fig2} plots the iterates of the first four elements of $\beta$ for one replicate of \lq P5-K1\rq{}
  example. Both MCMC methods seem to converge  quickly, with the  \optim{} method converging almost immediately and the \dr{} method also converging in less than 100 iterations. Figure  \ref{fig:fig3} is a similar plot for the \lq P10-K2\rq{}  example.
In general, the \optim{} method converges faster than the \dr{} method which takes around 250 iterations to converge.
We did not plot the iterates for $\btheta$, since they generally mix very well even in the higher dimensional \lq P10-K2\rq{}  example.
Figure \ref{fig:fig3} also plots the iterates when we set the parameter estimates from the \mcem{} method as initial values for the \dr{} method. In this case the chain converges almost immediately. In results for the \lq P10-F2\rq{}  example
that are not reported in the article, where a one-block strategy for sampling $\bbeta$ is used for \dr{}, it sometimes takes more than 1,000 iterations to obtain convergence which demonstrates the benefits of the sub-block strategy.

The results suggest that the estimates from the \mcem{} method can be used effectively as starting values for the MCMC methods because \mcem{} is much faster than either of the MCMC methods.

\subsection{Determining the number of factors} \label{ss: determing number of factors}
The \dr{} method is applied to fit the two \fmsv{} simulation examples with $k = 1, \dots, 3$ factors and evaluate their marginal likelihoods. We use the
Gaussian copula approximation method discussed in
section~\ref{s: marginal likelihood} to evaluate the
posterior ordinate $p(\btheta^*|\by,\bbeta^*)$ using one \lq Reduced MCMC\rq{}
simulation. A similar simulated example in \cite{nott:kohn:xu:fielding:2009}
 shows that the posterior ordinate $p(\btheta^*|\by,\bbeta^*)$ from a Gaussian
 copula approximation does  not change too much for different numbers of
 sub-block strategies. More importantly, it agrees with the evaluation from the
 benchmark joint kernel smoothing method using the many sub-blocks strategy
in \cite{chib:nardari:shephard:2006}.
Therefore, we adopt a one block strategy by only running one
\lq Reduced MCMC\rq{} simulation.   We run 5,000 iterations in sampling
$\btheta$ from $p(\btheta^*|\by,\bbeta^*)$, and 10,000 particles for $M$ and 20,000
particles for $R$ in the auxiliary particle filter used to calculate the likelihood.

Table \ref{tab:table4} reports the marginal likelihood evaluation results for
two simulated examples using  5 replicates. The true number of factors
in the \lq P5-K1\rq{}  and the \lq P10-K2\rq{}  cases are 1 and 2 respectively.
The marginal likelihoods suggest one (two) factors for the first (second) examples
for all 5 replicates. We did not evaluate the marginal likelihood
as in \cite{chib:nardari:shephard:2006}  because it is very slow due to
the multiple sub-blocks needed when applying kernel density estimation,
with the \lq optimization\rq{} method used for each `Reduced MCMC' sub-block (for evidence
see \cite{nott:kohn:xu:fielding:2009}). However, we believe that based on the
results of \cite{nott:kohn:xu:fielding:2009}, \dr{} combined with the
copula-based approximation method gives similar results to those
of \cite{chib:nardari:shephard:2006}, but is more practical  in terms of computational load.

\subsection{Faster computing speed}\label{ss: faster speed}
The computational burden for the \fmsv{} model usually becomes heavier
as  the dimension increases. However, given $\bB$ and $\bF$,
the model can be decomposed into several independent \usv{} models,
so that  parallel computing can be   used  in following areas.
\begin{enumerate}[I.]
\item Sampling $\bK_j, \btheta_j$ and $\bh_j$ for $(p+k)$ \usv{} equations
in both MCMC based methods. This can be applied in both the estimation stage
and later in the \lq Reduced MCMC\rq{}  needed to compute the
marginal likelihood.
\item	Calculating the gradient corresponding to
$\bbeta$ in the \fmsv{} model, which is needed
for all three methods discussed in section~\ref{Sec: factor msv models}.
\item	Sampling $\bK_i^j$ and $\bh_i^j$ for $i=1,...,p+k$, and evaluating the
complete log likelihood $p(\by,\bF^j,\bh^j,\bK^j|\bTheta)$ within each Monte
Carlo simulation $j=1,...,M,$ for the \mcem{} method.
\item	Calculating the importance weights in the auxiliary particle filter, which can be applied twice since the original auxiliary particle filter needs two re-sampling steps.
\end{enumerate}
\subsection{Computing details} \label{ss: computing details}
All the algorithms in the article  are coded in the Matlab M-language
running on a PC with Intel\textregistered Core 2 Quad CPU (3.0 GHz)
under the Matlab\textregistered 2009a framework with  the job assigned to 4 local workers.
We believe that the algorithms can be speeded up if coded in C, as in
\cite{chib:nardari:shephard:2006}.

\section{Real data application}
The \fmsv{} model is now fitted to a real dataset containing 18 international stock indices
that cover three major regions: America, Europe and Asian Pacific and
 both developed and emerging markets. Specifically, they are USA, Canada, Mexico, and
 Chile in America; UK, Germany, France, Switzerland, Spain, Italy, and Norway in Europe; and Japan, Hong Kong, Australia, New Zealand, Malaysia, Singapore, and Indonesia in Asia-Pacific. We use weekly continuously compounded returns (Wednesday to Wednesday) in nominal local currency from January 10th, 1990 to December 27th, 2006, giving a total of 886 weekly observations. The dataset is obtained from DataStream Morgan Stanley Capital International (MSCI) indices.

The \dr{} method with one iteration in the first stage is applied to estimate the model, using sub-blocks with 8 parameters in each sub-block, to sample $\beta$ from $p(\beta|y,\bh)$. We use marginal likelihood estimated using a Gaussian copula to select the number of factors, with two sub-blocks used to calculate the ordinate $p(\btheta^*|y,\beta^*)$.
Table~\ref{tab:table5} shows the log marginal likelihoods for $k=1,\dots, 5$ factors and
 suggests that the four factor model is the best with the Bayes factors of the 4-factor model compared with the other models  all greater than 100, providing strong evidence according to Jeffrey's scale. We do not list the parameter estimates due to space considerations, but the estimation results show good inefficiency levels: e.g. for the estimation of the 4-factor model we have average inefficiency factor scores
 of $\phi(13.2)$, $\sigma(20.3)$ and  $\beta(32.3)$, and acceptance ratios $(\phi,\sigma)(0.57)$ and $\beta(0.62)$ in the first stage.

\section{Other Applications}\label{e: other applications}
The \optim{} method discussed in \cite{chib:greenberg:1995}
is a powerful approach for MCMC simulation when it is necessary to form a Metropolis-Hastings proposal, either as a single block approach or using multiple sub-blocks. However, it can be quite slow. The \dr{} method discussed in our article can in principle be applied instead.  This section discusses several such applications.
\subsection{GARCH model}\label{ss: garch model}
This section applies the \dr{}  method to  the GARCH volatility model developed
by \cite{engle:1982} and generalized by \cite{bollerslev:1986}. We first consider the
univariate GARCH model which forms the basic building block of the factor multivariate
GARCH model discussed in section~\ref{ss: factor m garch}.

\subsubsection{Univariate model} \label{sss: univariate garch}
The popular univariate Gaussian-GARCH(1,1) model is
\begin{align}\label{eq: uni garch}
\begin{split}
y_t & =\epsilon_t, \hspace{0.5cm} \epsilon_t\sim N(0,\sigma^2_t)\ ,  \\
\sigma^2_t & =\omega+\alpha\epsilon^2_{t-1}+\beta\sigma^2_{t-1}.
\end{split}
\end{align}
Maximum likelihood estimation (MLE) is a convenient tool to estimate this  model,
but it may have trouble in more sophisticated GARCH type models such as the
factor Multivariate-GARCH (factor-MGARCH) model.
\cite{asai:2005} surveys work on Bayesian inference for the univariate
GARCH model and compares several existing estimation methods.

We  use two simulated examples to compare the performance of the \dr{} method
to the \lq Griddy Gibbs\rq{}  method  \citep[e.g.][]{bauwens:lubrano:1998}
the \optim{}  method, and  to the MLE.
Each example uses 1500 observations and 10 replications.
The true parameters  are $(\omega=0.1,\alpha=0.25,\beta=0.70)$ for example~1 and
$(\omega=0.1,\alpha=0.05,\beta=0.90)$ for example~2.
The volatility persistence parameter $\beta$ is moderate for example~1 and high for
example~2.

The priors used for the Bayesian methods are
\begin{align*}
\omega\sim \rmIG(2,\textrm{Var}(y)\times(1-0.95)), \hspace{0.5cm} \alpha\sim \textrm{Beta}(1,8), \hspace{0.5cm} \beta\sim \textrm{Beta}(8,1)\ ,
\end{align*}
where $\textrm{Var}(y)$ is the unconditional variance of $y$.
For the `Griddy-Gibbs' method we follow
\cite{bauwens:lubrano:1998}, and use  33 grid points with parameter ranges:
\begin{center}
$0<\omega<\textrm{Var}(y)\times(1-0.8), \hspace{0.5cm} 0<\alpha<0.3, \hspace{0.5cm} 0.35<\beta<1$\ .
\end{center}
Table \ref{tab:table6} reports the estimation results for the two simulated examples.
The parameter estimates are quite accurate for the\optim{} and \dr{}  and
 the MLE methods, but not for the `Griddy-Gibbs' method, whose parameter estimates
are far from the true values, especially in the second example where the
volatility persistence parameter is high. One reason for the poor performance
of  `Griddy-Gibbs' could be that the range of $\beta$ is too wide in this case,
and as suggested in \cite{bauwens:lubrano:1998}, we may further restrict its
range while allowing most of the posterior density to fit within this range.

The two block sampling MCMC methods (\optim{} and \dr{}) usually have lower
inefficiency factor scores than the single Gibbs based method, and  the \dr{} method
compares favourably with the \optim{}  method in terms of both inefficiency
factors and equivalence factor scores.

\subsubsection{Factor-MGARCH model}\label{ss: factor m garch}
Although  maximum likelihood estimation
is much faster than Bayesian methods in the univariate case,
Bayesian methods become more attractive for estimating factor
multivariate GARCH (factor-MGARCH)
models because it is much more difficult to apply the maximum likelihood estimation
in this case. See, for example, the discussion in \cite{harvey:ruiz:sentana:1992}
who need to make approximation in order to calculate the likelihood of a closely related
model. Under suitable assumptions on the \fmgarch{} model, parts of the computation can be
decomposed into working with several independent \ugarch{} models and the \dr{} can be
used to sample both the loading matrix and the parameters in each \ugarch{} model.

\subsection{Heavy tailed models}\label{ss: heavy tails}
 \cite{chib:nardari:shephard:2006} consider an \fmsv{} model with Student-t distributions for the idiosyncratic shocks, with $e_{j,t}\sim St(0,1,v_j)$ for $j=1,...,p$ in equation \eqref{eq:multi}, where $v_j$ is the degrees of freedom parameter in the Student-t distribution.

In a Bayesian framework, the Student-t error terms can be expressed in a conditionally normal form so that we  can use  \dr{} or the simpler \kstep{} method to build the proposal density in sampling the degrees of freedom parameters, instead of building the proposal densities using the \optim{} method. Similar ideas can  be applied in the t-GARCH model with heavy tails.


\section{Conclusions}\label{s: conclusion}
\Fmsv{} models provide a parsimonious representation of a dynamic multivariate system when
the time varying variances and covariances of the time series can be represented by a small
number of fundamental factors.
\Fmsv{} models can be used in many financial and economic application such as
portfolio allocation, asset pricing, and risk management.
MCMC methods are the main tool for estimating the parameters of \fmsv{} models and determining the number of factors in these models. In particular, the \optim{}
method  is widely used in the  literature to estimate
the parameters  of stochastic volatility models. Its main shortcoming is its computational expense, because numerical optimization  is required to build \MH{} proposals in each
MCMC iteration. The computational cost  is especially heavy for high dimensional
models containing many parameters.
Evaluating marginal likelihoods to determine the number of
factors is even more computationally expensive than model estimation because
it may require several reduced MCMC runs for each marginal likelihood evaluation.
We propose an alternative MCMC estimation method which is based on \dr{}
that is substantially faster than the \optim{} method and is more efficient
when measured in terms of equivalence factors.
We also propose a fast EM based approximation method for \fmsv{} models which can
be used either on its own or to provide initial parameter estimates to increase the speed of convergence of MCMC methods. We also note that we do not have a good way to determine the number of factors using the \mcem{} approach, whereas the Bayesian approach can use marginal likelihood.
Our article also simplifies the marginal likelihood calculation for determining the
number of latent factors in the \fmsv{} model by using Gaussian copula approximations
to the marginal ordinates instead of using kernel density estimates.
We also show how that our approach also applies in GARCH-type models  and compares
favorably with existing MCMC estimation methods.
The MCMC estimation methods and the fast EM based approximation method proposed in our article reduce the computational cost of estimating \fmsv{} models considerably, which is important for certain real-time applications in financial markets.

\bibliographystyle{asa}
\bibliography{factormsv}

\newpage

\begin{table}[t]
\begin{center}
    \footnotesize
    \begin{tabular}{crrrrrrrrrrrrrr}
    \hline\hline
    &  & \multicolumn{6}{c}{Sample Size T=500}  &     \multicolumn{6}{c}{Sample Size T=1,500 } \\ \cline{3-14}
    &  & Mean	& Stdev	& Ineff	& Equiv	& Acr	& Time  &   Mean	& Stdev	& Ineff	 & Equiv	 & Acr	& Time\\

   	& $\mu$   & 0.469 & 0.143 & 2.35  & 2.40  & 0.37  & 1.7(h) &  0.993 & 0.088 & 2.15  & 7.78  & 0.75  & 6.0(h) \\
   \raisebox{1.1ex}[0cm][-1cm]{Optimization-}
    & $\phi$  & 0.890 & 0.069 & 14.04 & 13.78 &       &       &   0.946 & 0.022 & 5.13  & 18.60 &       &  \\
   \raisebox{1.1ex}[0cm][-1cm]{MCMC-}
   	& $\sigma$& 0.117 & 0.035 & 8.18  & 8.33  &       &       &   0.130 & 0.028 & 6.12  & 22.19 &       &  \\
    &	&	&	&	&	&	&	&	&	&	&	& \\
\raisebox{1.1ex}[0cm][-1cm]{k-step}
    & $\mu$   & 0.471 & 0.143 & 2.33  & 1.05  & 0.33  & 0.8(h) &  0.994 & 0.088 & 2.14  & 3.02  & 0.67  & 2.4(h) \\
   \raisebox{1.1ex}[0cm][-1cm]{Iteration-}
    & $\phi$  & 0.897 & 0.066 & 15.77 & 7.15  &       &       &   0.947 & 0.021 & 7.37  & 10.41 &       &  \\
   \raisebox{1.1ex}[0cm][-1cm]{MCMC-}
   	& $\sigma$& 0.113 & 0.034 & 10.31 & 4.67  &       &       &   0.129 & 0.027 & 8.60  & 12.13 &       &  \\
    &	&	&	&	&	&	&	&	&	&	&	& \\

    Delayed-
    & $\mu$   & 0.469 & 0.159 & 2.07  & 1.12  & 0.32  & 0.9(h) &  0.993 & 0.088 & 2.10  & 3.23  & 0.67  & 2.6(h) \\
    Rejection-
 		& $\phi$  & 0.896 & 0.066 & 12.09 & 6.54  & 0.15  &       &   0.946 & 0.022 & 6.60  & 10.14 & 0.06  &  \\
    MCMC
	 	& $\sigma$& 0.115 & 0.035 & 8.08  & 4.34  &       &       &   0.130 & 0.028 & 8.00  & 12.27 &       &  \\
    &	&	&	&	&	&	&	&	&	&	&	& \\
		
		Monte
    & $\mu$   & 0.454 & 0.061 &       &       &       & 1.4(m) &  0.991 & 0.066 &       &       &       & 13.1(m) \\
    Carlo-
    & $\phi$  & 0.892 & 0.062 &       &       &       &       &   0.949 & 0.011 &       &       &  \\
    EM
   	& $\sigma$& 0.115 & 0.010 &       &       &       &       &   0.121 & 0.007 &       &       &  \\
	  \hline\hline
    \end{tabular}
\end{center}
\footnotesize
  \caption{{\bf Summary output of the univariate SV simulated examples}. The table reports the average values of posterior estimates and diagnostic measures across 10 replicates. `Mean', `Stdev', `Ineff', `Equiv', and `Acr' denote the posterior mean, posterior standard deviation, inefficiency factor, equivalence factor and acceptance ratio, respectively. Time is time in hours (h) and minutes (m). For \dr{} MCMC, there are two-stage acceptance ratios.}  	
  \label{tab:table1}
\end{table}

\begin{table}[htbp]
\begin{center}
    \begin{tabular}{cccccccccccc} %
    \hline\hline
          &		    & $y_1$    & $y_2$ & $y_3$    & $y_4$ & $y_5$ &       &       &       &       &         \\
P5-K1: Factor-Loadings $B$   & $f_1$ & $\bf{1}$ & -1.5  & 1.5   & -1.5  & 1.5   &       &       &       &       & \\
          &		    & $y_1$    & $y_2$ & $y_3$    & $y_4$ & $y_5$ & $y_6$ & $y_7$ & $y_8$ & $y_9$ & $y_{10}$  \\
          & $f_1$ & $\bf{1}$ &   0   &  0.5     &  -0.5 &  0.5  & -0.5  & 0.5   & -0.5  & 0.5   & -0.5 \\
\raisebox{1.1ex}[0cm][-1cm]{P10-K2: Factor-Loadings $B$} & $f_2$& $\bf{0}$ & $\bf{1}$ & 0.5  & -0.5 & 0.5  & -0.5 & 0.5  & -0.5 & 0.5  & -0.5 \\
    \hline\hline
    \end{tabular}
    \end{center}
    \footnotesize
    \caption{Values of the constrained (in bold) and unconstrained elements of the $\bB$ matrix}
    \label{T: B entries}
\end{table}
\begin{table}[t]
\begin{center}
    \footnotesize
    \begin{tabular}{crrrrrrrrcr}
    \hline\hline
    & \multicolumn{10}{c}{Panel A: Optimization-MCMC Method} \\ \cline{2-11}
    & $\mu$ & $\phi$ & $\sigma$ & $B$ &   & Ineff    & Equiv & \multicolumn{2}{c}{Acr} & Time \\

    $y_1$   & 0.555 & 0.884 & 0.121 &  \textbf{1.000}    & $\mu$    & 3.44  & 10.47 &       & $\left\{\phi,\sigma\right\}$ &  \\
    $y_2$   & 0.512 & 0.911 & 0.113 & -1.493  & $\phi$   & 15.99 & 48.71 & Stage(1) & 0.42  & 5.1(h) \\
    $y_3$   & 0.426 & 0.902 & 0.109 &  1.480  & $\sigma$ & 9.01  & 27.39 &       &       &  \\
    $y_4$   & 0.508 & 0.920 & 0.117 & -1.506  & $B$      & 7.02  & 21.40 &       & $B$   &  \\
    $y_5$   & 0.524 & 0.906 & 0.126 &  1.475  &       &       &       & Stage(1) & 0.87  &  \\
    $f_1$   & 1.102 & 0.968 & 0.123 &         &       &       &       &       &       &  \\ \hline

    &       &       &       &       &       &       &       &       &       &  \\
    & \multicolumn{10}{c}{Panel B: Delayed-Rejection: Iteration+Adaptive-MCMC Method} \\ \cline{2-11}

    $y_1$   & 0.556 & 0.896 & 0.117 &  \textbf{1.000}    & $\mu$    & 4.67  & 6.50  &       & $\left\{\phi,\sigma\right\}$ &  \\
    $y_2$   & 0.513 & 0.925 & 0.110 & -1.494  & $\phi$   & 14.68 & 20.37 & Stage(1) & 0.38  & 2.3(h) \\
    $y_3$   & 0.429 & 0.915 & 0.106 &  1.482  & $\sigma$ & 9.92  & 13.81 & Stage(2) & 0.11  &  \\
    $y_4$   & 0.505 & 0.929 & 0.115 & -1.503  & $B$      & 14.53 & 19.86 &       & $B$   &  \\
    $y_5$   & 0.521 & 0.912 & 0.126 &  1.473  &       &       &       & Stage(1) & 0.77  &  \\
    $f_1$   & 1.108 & 0.972 & 0.121 &         &       &       &       & Stage(2) & 0.01  &  \\ \hline
    &       &       &       &       &       &       &       &       &       &  \\

    & \multicolumn{10}{c}{Panel C: Monte Carlo EM Method} \\ \cline{2-11}
    $y_1$  & 0.539 & 0.881 & 0.099 &  \textbf{1.000}  &       &       &       &       &       & 5.9(m) \\
    $y_2$  & 0.495 & 0.891 & 0.098 & -1.450 &       &       &       &       &       &  \\
    $y_3$  & 0.418 & 0.878 & 0.096 &  1.435 &       &       &       &       &       &  \\
    $y_4$  & 0.472 & 0.901 & 0.099 & -1.463 &       &       &       &       &       &  \\
    $y_5$  & 0.516 & 0.900 & 0.101 &  1.432 &       &       &       &       &       &  \\
    $f_1$  & 1.116 & 0.953 & 0.131 &        &       &       &       &       &       &  \\
	  \hline\hline
    \end{tabular}
\end{center}
\footnotesize
  \caption{{\bf Summary output of \fmsv{} simulated examples for the
  `P5-K1' example.} In the table `P5-K1' means $p=5$ and $k=1$. The table reports the average values of the posterior estimates across 5 replicates. The inefficiency factor and equivalence factor scores are the average values for corresponding parameters across 5 replicates: e.g. the reported inefficiency factor for $\phi$  is the average value of inefficiency factors $(\phi_1,...,\phi_5)$ with each $\phi_j$ representing the average value of inefficiency factors $(\phi_{j,1},...,\phi_{j,p+k})$ for $j=1,...,5$.  The 5.9(m) computing time for \mcem{} does not include the cost of calculating the standard error.}
   \label{tab:table2}
\end{table}

\begin{table}[t]
\begin{center}
    \footnotesize
    \begin{tabular}{crrrrrrrrrcr}
    \hline\hline
    & \multicolumn{10}{c}{Panel A Optimization-MCMC Method} \\ \cline{2-12}
    & $\mu$ & $\phi$ & $\sigma$ & \multicolumn{2}{c}{$B$}  &  & Ineff  & Equiv & \multicolumn{2}{c}{Acr} & Time \\

    $y_1$  & 0.459 & 0.912 & 0.116 & \textbf{1.000}  & \textbf{1.000}    & $\mu$      & 10.43 & 145.27 &       & $\left\{\phi,\sigma\right\}$ &  \\
    $y_2$  & 0.436 & 0.922 & 0.125 &  0.003          & \textbf{1.000}  & $\phi$   & 16.30 & 225.61 & Stage(1) & 0.41  & 23.1(h) \\
    $y_3$  & 0.496 & 0.907 & 0.115 &  0.517 &  0.510                    & $\sigma$ & 9.67  & 133.73 &       &       &  \\
    $y_4$  & 0.573 & 0.892 & 0.110 & -0.526 & -0.515                    & $B$      & 41.74 & 583.17 &       & $B$   &  \\
    $y_5$  & 0.513 & 0.895 & 0.107 &  0.495 &  0.498 &       &       &                              & Stage(1) & 0.66  &  \\
    $y_6$  & 0.540 & 0.903 & 0.112 & -0.510 & -0.532 &       &       &       &       &       &  \\
    $y_7$  & 0.549 & 0.891 & 0.112 &  0.516 &  0.499 &       &       &       &       &       & \\
    $y_8$  & 0.533 & 0.893 & 0.122 & -0.507 & -0.526 &       &       &       &       &       &  \\
    $y_9$  & 0.496 & 0.913 & 0.114 &  0.515 &  0.497 &       &       &       &       &       &  \\
 $y_{10}$  & 0.541 & 0.902 & 0.112 & -0.515 & -0.484 &       &       &       &       &       &  \\
    $f_1$  & 1.131 & 0.967 & 0.127 &       &    &       &       &       &       &       &  \\
    $f_2$  & 0.956 & 0.957 & 0.134 &       &      &       &       &       &       &       &  \\  \hline
    &       &       &       &       &       &       &       &       &       &  \\

    & \multicolumn{10}{c}{Panel B Delayed-Rejection: Iteration+Adaptive-MCMC Method} \\ \cline{2-12}
    $y_1$  & 0.443 & 0.915 & 0.113 & \textbf{1.000} & \textbf{0.000}  & $\mu$     & 9.74  & 31.51 &       & $\left\{\phi,\sigma\right\}$ &  \\
    $y_2$  & 0.427 & 0.924 & 0.121 &  0.006         & \textbf{1.000}  & $\phi$    & 15.55 & 48.18 & Stage(1) & 0.37  & 4.4(h) \\
    $y_3$  & 0.494 & 0.915 & 0.113 &  0.506 &  0.524  & $\sigma$  & 10.42 & 33.25 & Stage(2) & 0.11  &  \\
    $y_4$  & 0.572 & 0.901 & 0.108 & -0.517 & -0.526  & $B$       & 40.10 & 131.90 &       & $B$ &  \\
    $y_5$  & 0.513 & 0.905 & 0.103 &  0.488 &  0.506  &       &       &       & Stage(1) & 0.72  &  \\
    $y_6$  & 0.539 & 0.910 & 0.109 & -0.499 & -0.545  &       &       &       & Stage(2) & 0.01  &  \\
    $y_7$  & 0.553 & 0.906 & 0.111 &  0.508 &  0.509  &       &       &       &       &       &  \\
    $y_8$  & 0.534 & 0.899 & 0.122 & -0.496 & -0.538  &       &       &       &       &       &  \\
    $y_9$  & 0.500 & 0.921 & 0.111 &  0.508 &  0.506  &       &       &       &       &       &  \\
 $y_{10}$  & 0.545 & 0.910 & 0.108 & -0.505 & -0.496  &       &       &       &       &       &  \\
    $f_1$  & 1.153 & 0.970 & 0.125 &       &  &       &       &       &       &       &  \\
    $f_2$  & 0.921 & 0.961 & 0.135 &       &      &       &       &       &       &       &  \\ \hline
    &       &       &       &       &       &       &       &       &       &  \\

    & \multicolumn{10}{c}{Panel C Monte Carlo EM Method} \\ \cline{2-12}
    $y_1$  & 0.501 & 0.881 & 0.098 & \textbf{1.000} & \textbf{0.000} &       &       &       &       &       & 8.8(m) \\
    $y_2$  & 0.462 & 0.906 & 0.100 & -0.013 & \textbf{1.000} &       &       &       &       &       &  \\
    $y_3$  & 0.466 & 0.903 & 0.098 & 0.500  & 0.497  &       &       &       &       &       &  \\
    $y_4$  & 0.536 & 0.861 & 0.097 & -0.507 & -0.505 &       &       &       &       &       &  \\
    $y_5$  & 0.493 & 0.865 & 0.096 & 0.483  & 0.479  &       &       &       &       &       &  \\
    $y_6$  & 0.526 & 0.878 & 0.097 & -0.493 & -0.514 &       &       &       &       &       &  \\
    $y_7$  & 0.538 & 0.870 & 0.097 & 0.499  & 0.489  &       &       &       &       &       &  \\
    $y_8$  & 0.523 & 0.888 & 0.098 & -0.486 & -0.514 &       &       &       &       &       &  \\
    $y_9$  & 0.465 & 0.904 & 0.098 & 0.500  & 0.485  &       &       &       &       &       &  \\
  $y_{10}$ & 0.540 & 0.872 & 0.097 & -0.501 & -0.475 &       &       &       &       &       &  \\
    $f_1$  & 1.169 & 0.959 & 0.132 &       &       &       &       &       &       &       &  \\
    $f_2$  & 0.888 & 0.949 & 0.134 &       &       &       &       &       &       &       &  \\
    \hline\hline
    \end{tabular}
\end{center}
\footnotesize
  \caption{{\bf Summary output of the \fmsv{}  `P10-K2' simulated example.} `P10-K2' means  $p=10$ and $k=2$. The table reports
  the average values of the posterior estimates across 5 replicates. The 8.8(m) computing time for \mcem{} does not include the cost of calculating the standard errors.}
   \label{tab:table3}
\end{table}

\begin{table}[t]
\begin{center}
    \footnotesize
    \begin{tabular}{crrrrrr}
    \hline\hline
    & \multicolumn{3}{c}{Simulation Example of $p=5$} &  \multicolumn{3}{c}{Simulation Example of $p=10$} \\ \cline{2-7}	
    & $k=1$(true) & $k=2$   & $k=3$   & $k=1$   & $k=2$(true) & $k=3$ \\
    Replicate-1 & {\bf -5,001.85} & -5,012.58 & -5,021.84 & -9,401.90 & {\bf -9,393.32} & -9,400.84 \\
    Replicate-2 & {\bf -4,967.76} & -4,974.77 & -4,983.24 & -9,563.01 & {\bf -9,551.47} & -9,558.68 \\
    Replicate-3 & {\bf -5,042.42} & -5,051.48 & -5,087.78 & -9,475.16 & {\bf -9,423.85} & -9,432.00 \\
    Replicate-4 & {\bf -5,060.63} & -5,069.60 & -5,078.10 & -9,427.84 & {\bf -9,407.43} & -9,418.41 \\
    Replicate-5 & {\bf -4,958.36} & -4,966.16 & -4,978.47 & -9,521.80 & {\bf -9,485.47} & -9,494.53 \\
    \hline\hline
    \end{tabular}
\end{center}
  \caption{Log Marginal likelihood estimates for the two \fmsv{} simulated examples}
   \label{tab:table4}
\end{table}

\begin{table}[t]
\begin{center}
    \footnotesize
    \begin{tabular}{cccccc}
    \hline\hline
    & \multicolumn{5}{c}{Sample from $01/1990\sim12/2006$} \\ \cline{2-6}	
    & $k=1$ & $k=2$   & $k=3$   & $k=4$   & $k=5$ \\
		Log Marginal-Likelihood & -34,097.05 & -33,722.09 & -33,610.93 & {\bf -33,554.69} & -33,594.61 \\
    \hline\hline
    \end{tabular}
\end{center}
  \caption{Log Marginal likelihood estimates for the real example}
   \label{tab:table5}
\end{table}

\begin{table}[t]
\begin{center}
    \footnotesize
    \begin{tabular}{crrrrrrrrrrrrrr}
    \hline\hline
    &  & \multicolumn{6}{c}{Simulation Example 1} &     \multicolumn{6}{c}{Simulation Example 2} \\ \cline{3-14}
    &  & Mean	& Stdev	& Ineff	& Equiv	& Acr	& Time  &   Mean	& Stdev	& Ineff	 & Equiv	 & Acr	& Time\\

        & $\omega$ & 0.102 & 0.022 &       &       &       & 1.2(s)&    0.107 & 0.058 &       &       &       & 1.4(s) \\
    MLE & $\alpha$ & 0.243 & 0.030 &       &       &       &       &    0.047 & 0.016 &       &       &       &  \\
        & $\beta$  & 0.703 & 0.032 &       &       &       &       &    0.899 & 0.040 &       &       &       &  \\
     &	&	&	&	&	&	&	&	&	&	&	& \\

        & $\omega$ & 0.077 & 0.031 & 6.65  & 3.95  &       & 59.5(m) & 0.227 & 0.119 & 30.12 & 17.60 &       & 58.4(m) \\
      \raisebox{1.1ex}[0cm][-1cm]{Griddy-}
        & $\alpha$ & 0.254 & 0.030 & 5.40  & 3.21  &       &       & 0.097 & 0.026 & 9.04  & 5.28  &       &  \\
      \raisebox{1.1ex}[0cm][-1cm]{Gibbs-}
   			& $\beta$  & 0.710 & 0.034 & 10.74 & 6.37  &       &       & 0.786 & 0.063 & 31.26 & 18.27 &       &  \\
     &	&	&	&	&	&	&	&	&	&	&	& \\

    		& $\omega$ & 0.095 & 0.017 & 3.17  & 1.30  & 0.58  & 40.6(m) & 0.101 & 0.029 & 5.16  & 2.22  & 0.46  & 41.9(m) \\
      \raisebox{1.1ex}[0cm][-1cm]{Optimization-}
		    & $\alpha$ & 0.231 & 0.029 & 8.43  & 4.02  &       &       & 0.048 & 0.013 & 5.68  & 2.44  &       &  \\
      \raisebox{1.1ex}[0cm][-1cm]{MCMC-}
  			& $\beta$  & 0.719 & 0.028 & 7.26  & 3.37  &       &       & 0.901 & 0.022 & 5.38  & 2.32  &       &  \\
     &	&	&	&	&	&	&	&	&	&	&	& \\

				Delayed-
    		& $\omega$ & 0.095 & 0.017 & 3.49  & 0.49  & 0.55  & 14.0(m) & 0.099 & 0.027 & 5.11  & 0.70  & 0.36  & 13.6(m) \\
				Rejection-
		    & $\alpha$ & 0.232 & 0.027 & 3.76  & 0.53  & 0.11  &       & 0.048 & 0.013 & 5.80  & 0.79  & 0.14  &  \\
   			MCMC
   			& $\beta$  & 0.718 & 0.027 & 4.10  & 0.58  &       &       & 0.903 & 0.021 & 5.61  & 0.76  &       &  \\
				\hline\hline
    \end{tabular}
\end{center}
\footnotesize
  \caption{{\bf Summary output of the univariate GARCH  simulated examples.} The table reports the average values of the posterior estimates across 10 replicates. For \dr{} MCMC, there are two-stage acceptance ratios.}    	
   \label{tab:table6}
\end{table}

\clearpage

\newpage

\end{document}